\documentclass[
aps,
physrev,
superscriptaddress,
amsmath,
amssymb,
nofootinbib,
floatfix,
twocolumn,
reprint
]
{revtex4-2}

\usepackage[utf8]{inputenc}
\usepackage{hyperref}
\usepackage{graphicx,color}
\usepackage{url}
\usepackage{bm}
\usepackage[italicdiff]{physics}
\usepackage{qcircuit}
\usepackage{amsthm}
\usepackage{enumitem}
\setlist[description]{leftmargin=\parindent,labelindent=\parindent}

\theoremstyle{plain}
\newtheorem{thm}{Theorem}
\newtheorem*{thm*}{Theorem}
\newtheorem*{prop*}{Proposition}

\newcommand{\tL}{\tilde{L}}
\newcommand{\mr}[1]{\mathrm{#1}}

\usepackage{hyperref}
\hypersetup{
colorlinks=true, 
linkcolor=blue, 
citecolor=blue 
}

\begin{document}

\title{Computation of Green’s function by local variational quantum compilation}
\author{Shota Kanasugi}
\affiliation{Quantum Laboratory, Fujitsu Research, Fujitsu Limited., 4-1-1 Kamikodanaka, Nakahara, Kawasaki, Kanagawa 211-8588, Japan}
\email{kanasugi.shota@fujitsu.com}

\author{Shoichiro Tsutsui}
\affiliation{QunaSys Inc., Aqua Hakusan Building 9F, 1-13-7 Hakusan, Bunkyo, Tokyo 113-0001, Japan}

\author{Yuya O. Nakagawa}
\affiliation{QunaSys Inc., Aqua Hakusan Building 9F, 1-13-7 Hakusan, Bunkyo, Tokyo 113-0001, Japan}

\author{Kazunori Maruyama}
\affiliation{Quantum Laboratory, Fujitsu Research, Fujitsu Limited., 4-1-1 Kamikodanaka, Nakahara, Kawasaki, Kanagawa 211-8588, Japan}

\author{Hirotaka Oshima}
\affiliation{Quantum Laboratory, Fujitsu Research, Fujitsu Limited., 4-1-1 Kamikodanaka, Nakahara, Kawasaki, Kanagawa 211-8588, Japan}

\author{Shintaro Sato}
\affiliation{Quantum Laboratory, Fujitsu Research, Fujitsu Limited., 4-1-1 Kamikodanaka, Nakahara, Kawasaki, Kanagawa 211-8588, Japan}

\date{\today}

\begin{abstract}
    Computation of the Green’s function is crucial to study the properties of quantum many-body systems such as strongly correlated systems. 
    Although the high-precision calculation of the Green's function is a notoriously challenging task on classical computers, the development of quantum computers may enable us to compute the Green's function with high accuracy even for classically-intractable large-scale systems. 
    Here, we propose an efficient method to compute the real-time Green's function based on the local variational quantum compilation (LVQC) algorithm, which simulates the time evolution of a large-scale quantum system using a low-depth quantum circuit constructed through optimization on a smaller-size subsystem. 
    Our method requires shallow quantum circuits to calculate the Green's function and can be utilized on both near-term noisy intermediate-scale and long-term fault-tolerant quantum computers depending on the computational resources we have. 
    We perform a numerical simulation of the Green's function for the one- and two-dimensional Fermi-Hubbard model up to $4\times4$ sites lattice (32 qubits) and demonstrate the validity of our protocol compared to a standard method based on the Trotter decomposition. 
    We finally present a detailed estimation of the gate count for the large-scale Fermi-Hubbard model, which also illustrates the advantage of our method over the Trotter decomposition.
\end{abstract}

\maketitle

\section{Introduction}\label{sec:Intro}
Simulation of quantum many-body systems is considered one of the most promising tasks for which quantum computers have a practical advantage over classical computers. 
Simulating quantum many-body systems is relevant to many scientific fields such as quantum chemistry, condensed matter physics, material science, and high-energy physics. 
Even noisy intermediate-scale quantum (NISQ) devices~\cite{preskill2018quantum} with a few hundred to thousands of qubits are believed to outperform classical computers in such quantum simulations, because of the exponential scaling of the Hilbert space with the size of the quantum system.
For instance, the variational quantum eigensolver (VQE) algorithm~\cite{peruzzo2014variational,kandala2017hardware,moll2018quantum,mcclean2016theory} enables us to compute the energy eigenvalues and eigenstates of quantum many-body systems on near-term NISQ devices. 

Another important quantity to study the nature of quantum many-body systems other than eigenvalues and eigenstates is the Green's function~\cite{bonch2015green,abrikosov2012methods,fetter2012quantum}. 
The Green's function provides us with much fundamental information about the quantum many-body systems. 
The Green's function is connected to various physical observables. 
For instance, the diagonal component of the Green's function is equal to the particle density of the system. 
The spectral function, which is directly calculated by the Fourier transform of the Green's function, gives the dispersion relation of quasiparticles. 
This is crucial information for studying strongly-correlated systems such as high-temperature superconductors~\cite{ding1996spectroscopic}. 
The Green's function also tells us the dynamical response of the quantum many-body systems under external perturbations, e.g., the linear response theory is described based on the Green's function~\cite{Kubo1957}. 

Many methods have been proposed in previous works to compute the Green's function on quantum computers. 
They are devised for either long-term fault-tolerant quantum computers (FTQCs) or near-term NISQ computers.  
For FTQCs, it has been proposed that the Green's function can be computed using quantum phase estimation~\cite{Wecker-PRA-2015,Roggero-PRC-2019,Kosugi-PRA-2020,Baker-PRA-2021}, Trotter decomposition of the time evolution operator~\cite{Bauer-PRX-2016,kreula2016non}, preconditioned linear system solver based on block encoding~\cite{QSVT-green-2021}, Gaussian integral transformation by qubitization technique~\cite{Roggero2020Gaussian}, and linear combination of unitary operations~\cite{keen2021LCU}. 
All of the above techniques generally require many high-fidelity qubits and gate operations and are hence suitable for long-term FTQCs. 
To circumvent such issues, various techniques that are suitable for computing Green's function on NISQ devices with limited hardware resources have been developed. 
These techniques include variational excited-states search methods~\cite{Endo-NISQ-2020,Zhu_2022}, generalized quantum equation of motion technique~\cite{qEOM-2022}, variational quantum simulation algorithm~\cite{Endo-NISQ-2020,Sakurai2022VQS,VQS-effective-2022,gomes2023computing}, combination of VQE and variational linear equation solver~\cite{VQEgreenPRA2021}, coupled cluster Green's function method~\cite{keen2022hybrid}, Krylov variational quantum algorithm~\cite{jamet2021krylov}, and Cartan decomposition~\cite{steckmann2021simulating}. 
These variational methods can be executed on NISQ devices in a quantum-classical hybrid manner. 
However, it is generally difficult to scale up these NISQ device techniques to large-scale quantum simulations. 
This issue could hinder NISQ devices from calculating the Green's function for practically important large-scale quantum many-body systems.  

\begin{figure*}[htbp]
    \includegraphics[keepaspectratio, scale=0.5]{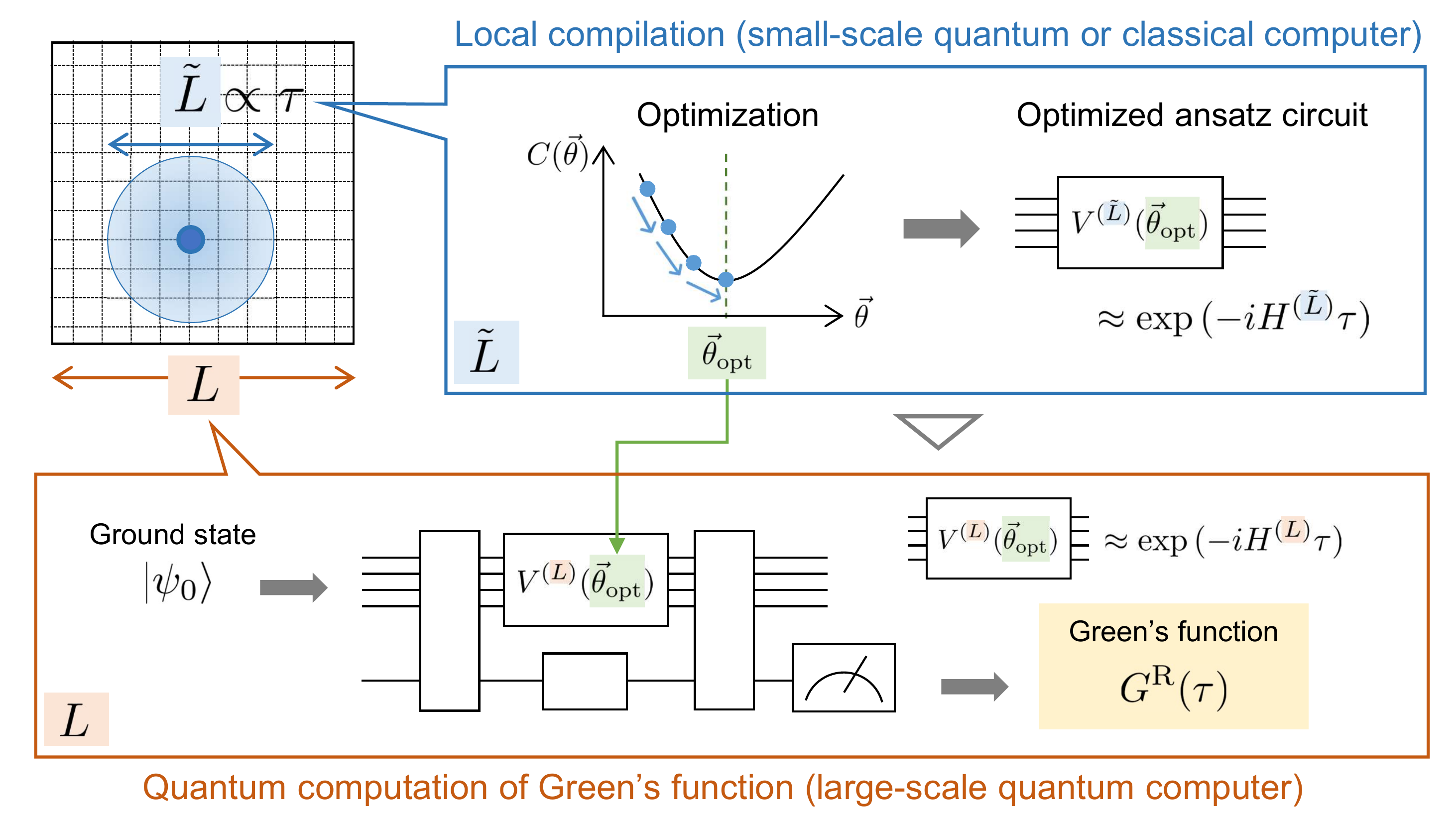}
    \caption{Overview of the LVQC approach to computing the Green's function of a quantum system with the lattice size $L$ at time $\tau$. First, we optimize the cost function $C(\vec{\theta})$ and find an optimal parameter $\vec{\theta}_{\rm opt}$ at a compilation size $\tilde{L}(< L)$ on small-scale quantum computers or classical computers. The compilation size $\tilde{L}\propto\tau$ is determined by the LR bound of the target system. The optimal parameter $\vec{\theta}_{\rm opt}$ gives an optimized circuit $V^{(\tilde{L})}(\vec{\theta}_{\rm opt})$ that approximates the time evolution operator at size $\tilde{L}$, i.e., $V^{(\tilde{L})}(\vec{\theta}_{\rm opt})\approx e^{-iH^{(\tilde{L})}\tau}$. 
    Next, we compute the Green's function on a large-scale quantum computer by using the optimal parameter $\vec{\theta}_{\rm opt}$ and the ground state $\ket{\psi_0}$. This process is executed by implementing an optimized circuit $V^{(L)}(\vec{\theta}_{\rm opt})$ that approximates the time evolution operator at size $L$, i.e., $V^{(L)}(\vec{\theta}_{\rm opt})\approx e^{-iH^{(L)}\tau}$. A long-time-scale dynamics at $t=n\tau$ ($n\in\mathbb{N}$) can also be calculated by adopting $(V^{(L)}(\vec{\theta}_{\rm opt}))^n\approx e^{-iH^{(L)}n\tau}$. } 
    \label{fig:lvqc-green}
\end{figure*}
In this paper, we propose a different approach to calculate the Green's function on quantum computers based on the local variational quantum compilation (LVQC) algorithm~\cite{Mizuta-LVQC}, which can bridge the gap between the methods for FTQCs and NISQ devices. 
The LVQC is a variational quantum algorithm to construct a low-depth quantum circuit that accurately approximates the time evolution operator for large-scale quantum many-body systems by optimizing the variational quantum circuit on a smaller-scale subsystem. 
Since the formulation of the LVQC relies mainly on the existence of the Lieb-Robinson (LR) bound~\cite{lieb1972finite}, it is applicable to the broad class of quantum many-body systems with local interactions. 
Specifically, the LVQC algorithm is implemented as follows. First, we execute the local compilation protocol, in which we optimize a low-depth variational quantum circuit to accurately approximate the time-evolution operator of a small subsystem. 
This optimization process may be executed by using NISQ devices and classical optimizers or only classical simulators. 
Then, we simulate the dynamics of a large-scale quantum system whose size lies in a regime intractable with the NISQ devices or the classical simulators. 
In this step, we use an optimized variational quantum circuit that approximates the time evolution operator of the large-scale quantum system, which is constructed by adopting the optimized parameter obtained in the local compilation process. 

Our proposal is to compute the Green's function of a large-scale quantum many-body system in the time domain by utilizing the approximate time evolution circuit constructed by the above LVQC protocol (Fig.~\ref{fig:lvqc-green}). 
The benefit of our LVQC-based method is that it is valid to reduce the circuit depth needed to accurately calculate the Green's function on a broad level of quantum computers from NISQ devices to FTQCs. 
Reducing the circuit depth is crucial for NISQ devices (and even for early FTQCs~\cite{PRXQuantum.3.010345,PRXQuantum.3.010318,kshirsagar2022proving,ding2022even,kuroiwa2023clifford+,Akahoshi-preFTQC}) to complete the computation within the coherence time and to alleviate the accumulation of gate error. 
The reduction of the circuit depth is also essential for ideal FTQCs to reduce the total simulation time.
We show the validity of our method by performing numerical simulations of the Fermi-Hubbard model~\cite{arovas2022hubbard}, which is the simplest model of interacting fermions but is essential to study the nature of strongly correlated electron systems. 
We also estimate the gate count for quantum circuits to calculate the Green's function for the large-scale Fermi-Hubbard model and illustrate the reduction of the gate count in our method.
Although we focus on the Fermi-Hubbard model in this paper, our LVQC-based method is applicable to compute the Green's function for a variety of quantum many-body Hamiltonian with the LR bound. 

The rest of this paper is organized as follows. 
In Sec.~\ref{sec:Green}, we summarize the definition of the Green's function and a method to calculate the Green's function on a quantum computer. 
In Sec.~\ref{sec:LVQC}, we briefly review the LVQC algorithm. 
In Sec.~\ref{sec:protocol}, we propose a protocol to calculate the Green's function based on the LVQC algorithm.
In Sec.~\ref{sec:Numerical}, we demonstrate the validity of the LVQC algorithm for computing the Green's function by performing numerical simulations of the one- and two-dimensional Fermi-Hubbard model. 
In Sec.~\ref{sec:ResourceEstimation}, we estimate the computational resources needed to apply the LVQC algorithm in large quantum systems.
Finally, we discuss some remarks on our results and conclude this paper in Sec.~\ref{sec:Summary}. 

\section{Review of Green's function and its calculation on quantum computers }\label{sec:Green}
In this section, we review the definition of quantities we focus on in this study, namely, the Green's function, the spectral function, and the density of states (DOS).
We then explain a general strategy to calculate them on quantum computers based on the decomposition of the Green's function into the sum of outputs of specific quantum circuits, which was employed in various studies~\cite{Bauer-PRX-2016,kreula2016non,Endo-NISQ-2020,VQS-effective-2022}.

\subsection{Definition of Green's function and related physical quantities}
For a fermionic system described by Hamiltonian $H$, the retarded Green's function at zero temperature is defined as 
\begin{equation}
    G^{\rm R}_{a,b}(t) = -i\Theta(t)\bra{\psi_0}\{e^{iHt}c_{a}e^{-iHt}, c_{b}^{\dag}\}\ket{\psi_0},
    \label{eq:def-green}
\end{equation}
where $\{A,B\}=AB+BA$ denotes the anticommutator, $\Theta(t)$ is the Heaviside step function, $\ket{\psi_0}$ is the ground state of Hamiltonian, and $c_a$ and $c_{b}^{\dag}$ are fermionic annihilation and creation operators, respectively.
The index $a \: (=1,2,\cdots,M)$ of the operators $c_a$, $c_a^{\dag}$ specifies the fermionic mode, where $M$ denotes the total number of fermionic modes. 
To discuss the spectral function, we here set $a=\mathbf{x}\sigma$, where $\mathbf{x}$ denotes the spatial coordinate and $\sigma=\uparrow,\downarrow$ denotes the spin of the fermionic particle. 
Hereafter, we consider only the spin-diagonal component of the Green's function for simplicity. 
Then, the Green's function in the momentum space is defined as
\begin{equation}
    G^{\rm R}_{\mathbf{k}\sigma}(t) = \frac{1}{V}\sum_{\mathbf{x},\mathbf{x}'}e^{-i\mathbf{k}\cdot(\mathbf{x}-\mathbf{x}')}G^{\rm R}_{\mathbf{x}\sigma,\mathbf{x}'\sigma}(t),\label{eq:gf-momentum}
\end{equation}
where $\mathbf{k}$ denotes the momentum and $V$ denotes the volume of the system. 
Using $G^{\rm R}_{\mathbf{k}\sigma}(t)$, the spectral function for fermions of momentum $\mathbf{k}$ and spin $\sigma$ is obtained as 
\begin{equation}
    A_{\mathbf{k}\sigma}(\omega) = -\frac{1}{\pi}\mathrm{Im} \, \tilde{G}^{\rm R}_{\mathbf{k}\sigma}(\omega).
    \label{eq:def-spectral}
\end{equation}
Here, $\tilde{G}^{\rm R}_{\mathbf{k}\sigma}(\omega)$ is the Fourier transform of $G^{\rm R}_{\mathbf{k}\sigma}(t)$,
\begin{equation}
    \tilde{G}^{\rm R}_{\mathbf{k}\sigma}(\omega) = \int_{-\infty}^{\infty}dt e^{i(\omega+i\eta)t}G^{\rm R}_{\mathbf{k}\sigma}(t), \label{eq:green-fourier}
\end{equation}
where $\eta(\to+0)$ is an infinitesimal positive number to ensure the convergence of the integral. 
The spectral function is a fundamental quantity in quantum many-body physics, which contains information about the energy distribution in the momentum space~\cite{bonch2015green,abrikosov2012methods,fetter2012quantum}. 
More specifically, the peak location of the spectral function in the $(\mathbf{k},\omega)$-space corresponds to the dispersion relation of the quasiparticles and the height of the peak describes the probability of finding a quasiparticle with a specific momentum and energy.  
The spectral function is related to the DOS as
\begin{equation}
    \rho_{\sigma}(\omega) = \frac{1}{V}\sum_{\mathbf{k}}A_{\mathbf{k}\sigma}(\omega).
    \label{eq:dos}
\end{equation}
The DOS describes the number of energy states per unit volume that can be occupied by electrons. 
The DOS can be used to estimate the band gap of solid materials, which is crucial information for studying semiconductors and superconductors. 
In addition, many bulk properties of solid materials (e.g., specific heats, magnetic susceptibility, and conductivity) are often described by the DOS.  
Both the spectral function and DOS provide us with a lot of crucial information about the electronic structure of materials and can be measured using spectroscopic techniques e.g., angle-resolved photoemission spectroscopy~\cite{ARPES-review} and scanning tunneling spectroscopy~\cite{STS-STM-review}.

\subsection{Quantum computation of real-time Green's function}\label{subsec:Green-quantum}
\begin{figure}[tbp]
    \includegraphics[keepaspectratio, scale=0.92]{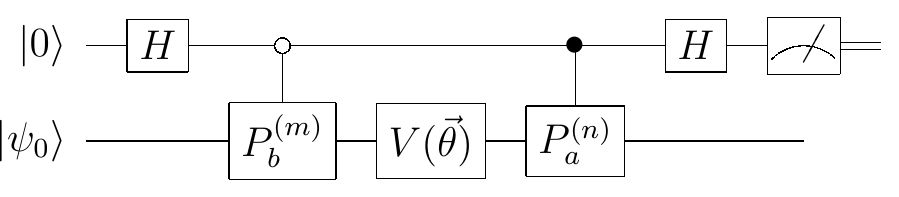}
    \caption{Quantum circuit to compute $K_{a,b}^{(n,m)}(t)$ given by Eq.~\eqref{eq:gf-pauli-kernel}. The upper line represents the ancillary qubit and the lower line represents the qubits for the system of interest. The expectation value of $Z$ measurement on the ancillary qubit yields $\mathrm{Re}\bra{\psi_0}V^{\dag}(\vec{\theta})P_{a}^{(n)}V(\vec{\theta})P_{b}^{(m)}\ket{\psi_0}$, which is an approximate value of $K_{a,b}^{(n,m)}(t)$ when $V(\vec{\theta})\approx e^{-iHt}$. }
\label{fig:circuit-gf}
\end{figure}

Here, we explain a general way to evaluate the Green's function on a quantum computer, on which our propsal in Sec.~\ref{sec:protocol} is also based. 
The first step is to prepare the (approximate) ground state $\ket{\psi_0}$ for a given Hamiltonian $H$. 
We assume that this task can be accomplished with high accuracy by utilizing quantum algorithms such as VQE~\cite{peruzzo2014variational,kandala2017hardware,moll2018quantum,mcclean2016theory} on near-term NISQ devices or quantum phase estimation~\cite{kitaev1995quantum,cleve1998quantum,nielsen-chuang} on long-term FTQCs.  
Although it may be possible to directly compute Green's function using long-term algorithms (e.g., quantum phase estimation~\cite{Wecker-PRA-2015,Roggero-PRC-2019,Kosugi-PRA-2020,Baker-PRA-2021} or Trotter decomposition~\cite{Bauer-PRX-2016,kreula2016non}) if we can prepare the ground state by quantum phase estimation, our LVQC-based method will be still useful to reduce the circuit depth and total simulation time.
Next, we decompose the fermionic operators $c_a$, $c_a^{\dag}$ into a sum of Pauli matrices as, 
\begin{align}
  \begin{split}
    c_{a} \mapsto \sum_{n}\lambda_{a}^{(n)}P_{a}^{(n)}, \quad 
    c_{a}^{\dag} \mapsto \sum_{n}\lambda_{a}^{(n)*}P_{a}^{(n)},\label{eq:qubit-map}
  \end{split}
\end{align}
where $P_{a}^{(n)}$ is tensor product of Pauli matrices. 
We can generally realize the decomposition~\eqref{eq:qubit-map} by adopting bosonization techniques such as Jordan-Wigner encoding~\cite{Jordan1928}, Bravyi-Kitaev encoding~\cite{bravyi-kitaev-2002} and parity encoding~\cite{seeley2012bravyi}. 
From Eq.~\eqref{eq:qubit-map}, the retarded Green's function~\eqref{eq:def-green} can be rewritten as 
\begin{equation}
    G_{a,b}^{\rm R}(t) = -2i\Theta(t)\sum_{n,m}\lambda_{a}^{(n)}\lambda_{b}^{(m)*} K_{a,b}^{(n,m)}(t),
    \label{eq:gf-pauli}
\end{equation}
where $K_{a,b}^{(n,m)}(t)$ is defined as 
\begin{equation}
    K_{a,b}^{(n,m)}(t) = \mathrm{Re}\bra{\psi_0}e^{iHt}P_{a}^{(n)}e^{-iHt}P_{b}^{(m)}\ket{\psi_0}.
    \label{eq:gf-pauli-kernel}
\end{equation}
The problem is now reduced to evaluate the value of $K_{a,b}^{(n,m)}(t)$ on a quantum computer.
Since the exact time-evolution operator $e^{-iHt}$ cannot be implemented on quantum computers in general, we need to approximate $e^{-iHt}$ in terms of native quantum gates. 
If we have such a quantum circuit $V(\vec{\theta})$ that approximates the time-evolution operator $e^{-iHt}$, we can compute the Green's function by utilizing the quantum circuit shown in Fig.~\ref{fig:circuit-gf}. 
The repeated measurement of the ancillary qubit yields the expectation value $\mathrm{Re}\bra{\psi_0}V^{\dag}(\vec{\theta})P_{a}^{(n)}V(\vec{\theta})P_{b}^{(m)}\ket{\psi_0}$, which is approximately equal to $K_{a,b}^{(n,m)}(t)$ when $V(\vec{\theta})\approx e^{-iHt}$. 
To obtain all components of the Green's function, such measurements should be performed for $N_{\rm circ}$ distinct quantum circuits, where $N_{\rm circ}$ is the number of all possible patterns of $K_{a,b}^{(n,m)}$ with respect to $(a,b; n,m)$. 
If we choose a fermion-to-qubit mapping that represents $c_a$ and $c_a^{\dag}$ as linear combinations of two independent Pauli operators (see Appendix~\ref{append:Majorana}), we obtain $N_{\rm circ}=4M^2$ that can be very large in large-scale quantum systems.
Fortunately, such difficulty can be mitigated at some level by considering the symmetry of the target Hamiltonian $H$ (see Appendix~\ref{append:symmetry}).

\section{Local variational quantum compilation on fermionic systems} \label{sec:LVQC}
In this section, we explain LVQC proposed in Ref.~\cite{Mizuta-LVQC}.
LVQC enables us to determine the variational quantum circuit that approximates the time evolution operator for the whole system, $V(\vec{\theta})\approx e^{-iHt}$, by using only the smaller subsystem(s).
Our strategy to calculate the Green's function, which will be explained in the next section, is to leverage the approximate time evolution operator obtained by LVQC.
We note that although LVQC was already proposed in Ref.~\cite{Mizuta-LVQC} for general systems, its formulation was explained only for spin systems.
Our contribution in this section is to show explicit procedures of LVQC for fermionic systems (e.g., how to take the subsystems).

\subsection{Setup}
We consider a lattice $\Lambda = \{1,2,\cdots, L\}$, or a set of indices of ``sites", and define two fermionic modes corresponding to spin-up and spin-down ($\sigma=\uparrow,\downarrow$) on each site.
The annihilation (creation) operators on the site $i$ with spin $\sigma$ is denoted by $c_{i\sigma} (c_{i\sigma}^\dag)$.
For simplicity, we assume the one-dimensional lattice with the periodic boundary condition and the Hamiltonian is translationally invariant:
\begin{equation} \label{eq: def H_N_PBC}
 H^{(L)}_{f, \mr{PBC}} = \sum_{i=1}^L h_{i,i+1}, \:\: h_{i,i+1} = \mathcal{T}^{i-1} (h_{1,2}),
\end{equation}
where $\mathcal{T}$ is a one-site translation operator $\mathcal{T}(c_{i\sigma}) = c_{i+1 \sigma}$ with identifying $i=L+1$ as $i=1$ and $h_{1,2}$ consists of even numbers of the annihilation and creation operators on the sites $1$ and $2$.
We note that $h_{i,j}$ conserves the parity of the number of fermions and commutes with $h_{k,l}$ if $i,j,k,l$ are mutually different. 
Extensions to other lattices, dimensions, the range of the interactions (as far as it is finite), and not-translationally-invariant cases can be performed straightforwardly (see also Appendix~\ref{append:LVQC-fermion}).

We also consider a local subsystem $\Lambda^{(j, \tL)}$ consisting of $\tL$ sites centered at the site~$j$,
\begin{equation}
 \Lambda^{(j, \tL)} = \{ j - \tL/2 + 1, \cdots, j + \tL/2 \}.
 \label{eq: def of lamnda_j_tN}
\end{equation}
Since we assume the translation invariance, we take $j=\tL/2$ and denote
\begin{equation}
 \Lambda^{(\tL)} := \Lambda^{(\tL/2, \tL)} = \{ 1, 2, \cdots, \tL \}.
 \label{eq: def of lamnda_tN}
\end{equation}
The local subsystem Hamiltonian with the periodic boundary condition for $\Lambda^{(\tL)}$ is defined as
\begin{equation} \label{eq: def H_tildeN_PBC}
 H^{(\tL)}_{f,\mr{PBC}} = \sum_{i=1}^{\tL} \tilde{h}_{i,i+1}, \:\: \tilde{h}_{i,i+1} = \tilde{\mathcal{T}}^{i-1}(h_{1,2}),
 \end{equation}
 where the translation operator $\tilde{\mathcal{T}}$ acts as $\tilde{\mathcal{T}} (c_{i\sigma}) = c_{i+1 \sigma}$ with identifying $i=\tL+1$ as $i=1$.
 We note that $ H^{(\tL)}_{f,\mr{PBC}}$ is the same as the restriction of $ H^{(L)}_{f,\mr{PBC}}$ on $\Lambda^{(\tL)}$ expect for the boundary term $\tilde{h}_{\tL, \tL+1}$.

To approximate the time evolution operator of the total system $U_f^{(L)}(\tau) := e^{- i H_{f,\mr{PBC}}^{(L)} \tau}$ for a fixed time $\tau$, we consider the (translationally-invariant) ansatz $V^{(L)}(\vec{\theta})$ of the depth $d$ in the brick-wall structure,
\begin{equation} \label{eq: ansatz for N}
 V_f^{(L)}(\vec{\theta}) := \prod_{k=1}^d \left[ \left( \prod_{i=1}^{L/2} V^{(k)}_{2i,2i+1}(\theta_{2k+1}) \right) \left( \prod_{i=1}^{L/2} V^{(k)}_{2i-1,2i}(\theta_{2k}) \right) \right],
\end{equation}
where $V^{(k)}_{i,i+1}(\theta)$ acts nontrivially only on the sites $i$ and $i+1$, conserves the parity of the number of fermions, and is translationally invariant with respect to $i$ by identifying $i=L+1$ as $i=1$ (i.e., $V^{(k)}_{i,i+1}(\theta) = \mathcal{T}^{i-1} (V^{(k)}_{1,2}(\theta))$).
For example, $V^{(k)}_{i,i+1}(\theta)$ can be a fermionic rotational gate like $e^{i\theta  \sum_\sigma (c_{i\sigma}^\dag c_{i+1\sigma} - c_{i+1\sigma}^\dag c_{i\sigma})}$.
The number of the parameters $\vec{\theta}$ is $2d$, $\vec{\theta}=(\theta_1, \cdots, \theta_{2d})$.
Similarly, to approximate the time evolution operator of the local Hamiltonian,
$U_f^{(\tL)}(\tau) = e^{-i H^{(\tL)}_{f, \mr{PBC}} \tau}$, the local version of the ansatz is defined as,
\begin{equation} \label{eq: ansatz for tildeN}
 V_f^{(\tL)}(\vec{\theta}) = \prod_{k=1}^d \left[ \left( \prod_{i=1}^{\tL/2} \tilde{V}^{(k)}_{2i,2i+1}(\theta_{2k+1}) \right) \left( \prod_{i=1}^{\tL/2}\tilde{V}^{(k)}_{2i-1,2i}(\theta_{2k}) \right) \right],
\end{equation}
where $\tilde{V}^{(k)}_{i,i+1}(\theta)$ is translationally invariant with respect to $i$ by identifying $i=\tL+1$ as $i=1$, i.e., $\tilde{V}^{(k)}_{i,i+1}(\theta) = \tilde{\mathcal{T}}^{i-1} (V^{(k)}_{1,2}(\theta))$.

\subsection{LVQC cost functions}
The LVQC algorithm aims at approximating the time-evolution operator of the $L$-sites system $U_f^{(L)}(\tau)$ by the variational quantum circuit $V_f^{(L)}(\vec{\theta})$.
Here we introduce two cost functions that measure the distance between two unitaries $U_f$ and $V_f$ written in fermion operators. 

Let us consider the Hilbert space generated by $M$ fermionic mode, i.e., $\mathbb{H}_M = \{ \ket{n_1, \cdots, n_M}_f = (c_1^\dag)^{n_M} \cdots, (c_M^\dag)^{n_M} \ket{\mr{vac}} \mid n_\mu = 0,1 \, (\mu=1,\cdots,M) \}$, where $\ket{\mr{vac}}$ is the vacuum state.
In our case of the spinful fermions on the lattice $\Lambda$, $\mu$ is a tuple of the site and spin, $(i, \sigma)$, and $M=2L$.
For two unitaries $U_f, V_f$ on $\mathbb{H}_M$, we define the Hilbert-Schmidt test (HST) cost function
\begin{equation}
 C_\mr{HST}^f (U_f, V_f) = 1 - \frac{1}{4^M} \left|\Tr (U_f V_f^\dag) \right|^2.
\end{equation}
This cost function has two properties,
(1) $0 \leq C_\mr{HST}^f (U_f, V_f) \leq 1$ (positiveness),
(2) $C_\mr{HST}^f (U_f, V_f) = 0 \Leftrightarrow U_f = e^{i\phi} V_f$ for some $\phi \in \mathbb{R}$ (faithfullness).
Therefore, one can use this cost function to find the approximation of $U_f$ by minimizing it with varying $V_f$.

We can rewrite the HST function by using the ``fermion Bell pair" between a doubled system, which consists of two identical Hilbert spaces $\mathbb{H}_M$ named $A$ and $B$. 
The fermion Bell pair between the system $A$ and $B$ is defined as
\begin{align}
\ket{\Phi^f_{+,\mu}} &:= \frac{1}{\sqrt{2}} (\ket{\mr{vac}}_{A_\mu B_\mu} + c_{A_\mu}^\dag c_{B_\mu}^\dag\ket{\mr{vac}}_{A_\mu B_\mu}) \label{eq: def bell pair mu} \\
\ket{\Phi^f_+} &:= \bigotimes_{\mu=1}^M \ket{\Phi^f_{+,\mu}},  \label{eq: def bell pair}
\end{align}
where $A_\mu (B_\mu)$ represents the $\mu$-th mode of the system $A(B)$.
We can show 
\begin{equation}
 C_\mr{HST}^f(U_f,V_f)= 1 - \left| \bra{\Phi_{+}^f} (U_f \otimes V_f^*) \ket{\Phi_{+}^f} \right|^2.
\end{equation}

Inspired by this expression, the second cost function we consider is the local Hilbert-Schmidt test (LHST) cost function, defined as 
\begin{equation}
 C_\mr{LHST}^f(U_f,V_f) = \frac{1}{M} \sum_{\mu=1}^{M} C_\mr{LHST}^{(\mu),f}(U_f,V_f), 
\end{equation}
where
\begin{equation} \label{eq: def C_LHST_mu}
 C_{\rm LHST}^{(\mu),f}(U_f,V_f) = 1 - \left|\bra{\Phi^f_{+,\mu}} (U_f \otimes V_f^*) \ket{\Phi^f_+} \right|^2. 
\end{equation}
$C_\mr{LHST}^{(\mu),f}(U_f,V_f)$ has the positiveness and faithfulness: $0 \leq C_\mr{LHST}^{(\mu),f}(U_f,V_f) \leq 1$ and $ C_\mr{LHST}^{(\mu), f}(U_f,V_f) = 0 \Leftrightarrow U_f V_f^\dag = e^{i\phi} I_{\mu} \otimes W$, where $I_\mu$ is the identity operator for the mode $\mu$, $W$ is a unitary acting on the modes $1, \cdots, \mu-1, \mu+1, \cdots, M$, and $\phi$ is some real number.
These properties result in the positiveness and faithfulness of the LHST cost function,
$0 \leq C_\mr{LHST}^f(U_f,V_f) \leq 1$ and
$C_\mr{LHST}^f (U_f, V_f) = 0 \Leftrightarrow U_f = e^{i\phi} V_f$ for some $\phi \in \mathbb{R}$.

\subsection{LVQC algorithm}
Roughly speaking, LVQC states that the optimization of the local ansatz $V^{(\tL)}(\vec{\theta})$ to the local time evolution operator $U_f^{(\tL)}(\tau)$ in the local system of $\tL$ sites suffices to find the parameters $\vec{\theta}_\mr{opt}$ that approximates the time evolution in the total system of $L$ sites: $U_f^{(L)}(\tau) \approx V_f^{(L)}(\vec{\theta}_\mr{opt})$.
The physics behind LVQC is the existence of the Lieb-Robinson (LR) bound~\cite{lieb1972finite}, which dictates that any local observable cannot spread out faster than a certain velocity under a local Hamiltonian. 
For a local Hamiltonian $H$ on the lattice $\Lambda$ such as $H_{f,\mr{PBC}}^{(L)}$ and local observables $O_X, O_Y$ which consists of the even number of fermionic operators and acts nontrivially on the domains $X,Y \subseteq{\Lambda}$ with normalization $\| O_X \| = \| O_Y \| = 1$, the LR bound is expressed as follows:
\begin{equation}
    \left\| \left[e^{iH\tau} O_{X} e^{-iH\tau},O_{Y} \right] \right\| \leq Ce^{-[\mathrm{dist}(X,Y)-v\tau]/\xi},
    \label{eq:LR-bound}
\end{equation}
where $[P,Q]=PQ-QP$ is the commutator, $\tau$ is a fixed time, $\mathrm{dist}(X,Y)$ is the distance between the domains, and $\|\cdot\|$ denotes the operator norm. 
The LR velocity $v$, the length $\xi$, and the coefficient $C$ are constants with respect to $L$, determined solely by the property of the local Hamiltonian $H$ such as the range of the interactions.
The velocity $v$ and the length $\xi$ do not depend on $\tau$ while $C$ typically depends on $\tau$~\cite{Mizuta-LVQC}. 

For the translationally-invariant Hamiltonian on the one-dimensional lattice defined above, a theorem of LVQC can be stated as follows:
\begin{thm}[LVQC for translationally-invariant fermionic systems] \label{thm:LVQC}
Consider fermionic systems of $L$ sites and $\tL$ sites and the Hamiltonians on them, $H^{(L)}_{f,\mr{PBC}}$ and $H^{(\tL)}_{f,\mr{PBC}}$ respectively [Eqs.~\eqref{eq: def H_N_PBC} and~\eqref{eq: def H_tildeN_PBC}].
Assume that the ansatzes on the two systems has the form of Eqs.~\eqref{eq: ansatz for N} and \eqref{eq: ansatz for tildeN} with depth $d$.
We choose the compilation size $\tL$ as
\begin{equation} 
    \tL \geq l_0+d_{H}+v\tau + 2d + 1, \label{eq:compile-size}
\end{equation}
where $l_0$ is a tunable parameter determining $\tL$, $d_H=\order{L^0}=\order{1}$ is the range of interaction, $\tau$ is a fixed time, and $v$ is the velocity of the system in the LR bound~\eqref{eq:LR-bound}.
Suppose that we find optimal parameters $\vec{\theta}_{\rm opt}$ by minimizing the HST or LHST cost functions on the local system of $\tL$ sites,  which satisfy
\begin{align}
    C^f_{\rm LHST}( U^{(\tL)}_{f}(\tau),V_f^{(\tL)}(\vec{\theta}_{\rm opt})) &< \varepsilon_{\rm LHST}, \\
    C^f_{\rm HST}( U^{(\tL)}_{f}(\tau), V_f^{(\tL)}(\vec{\theta}_{\rm opt})) &< \varepsilon_{\rm HST}, 
\end{align}
for some $\varepsilon_\mr{LHST} > 0$ and $\varepsilon_\mr{HST} > 0$.
The following equations hold for the total system of $L$ sites,
\begin{align}
 C^f_{\rm LHST}( U^{(L)}_{f}(\tau), V_f^{(L)}(\vec{\theta}_{\rm opt})) &\leq \varepsilon_{\rm LHST} + 4\varepsilon_{\rm LR}, \label{eq:cost-lhst-lvqc} \\
 C^f_{\rm HST}(  U^{(L)}_{f}(\tau), V_f^{(L)}(\vec{\theta}_{\rm opt})) &\leq 2L \left(\varepsilon_{\rm HST} + 4 \varepsilon_{\rm LR}\right), \label{eq:cost-hst-lvqc}
\end{align}
where $\varepsilon_{\rm LR}=e^{-\mathcal{O}(l_0/\xi)}$ with $\xi$ being the length scale appearing in the LR bound [Eq.~\eqref{eq:LR-bound}]. 
\end{thm}
\noindent
This theorem indicates that we can employ the optimal parameter set $\vec{\theta}_{\rm opt}$ for the size-$\tL$ system to approximate the time evolution operator of the size-$L$ system as $U_f^{(L)}(\tau) = e^{-iH^{(L)}_{f,\mr{PBC}} \tau}\approx V_f^{(L)}(\vec{\theta}_{\rm opt})$ (recall the faithfulness of $C_\mr{LHST}^f$ and $C_\mr{HST}^f$).
The proof of LVQC theorem was presented in Ref.~\cite{Mizuta-LVQC} for general spin systems, but the extension to fermionic systems is not so different since the proof depended mostly on the existence of the LR bound and not on the nature of the particles (spins, bosons, fermions). 
For completeness, we describe the sketch of the proof and specific remarks on the LVQC theorem for fermionic systems in Appendix~\ref{append:LVQC-fermion}.

From Eqs.~\eqref{eq:cost-lhst-lvqc} and~\eqref{eq:cost-hst-lvqc}, we see that the error of the LVQC protocol consists of two parts. 
The first one stems from the limitation of the expressive power of the ansatz $V^{(\tL)}$, denoted by $\varepsilon_{\rm LHST}$ and $\varepsilon_{\rm HST}$. 
This error can be improved by using, for example, a more expressive ansatz, appropriate initial parameters, and sophisticated classical optimizers in the optimization. 
The second one is an intrinsic error owing to the nature of the LR bound, denoted by $\varepsilon_{\rm LR}=e^{-\mathcal{O}(l_0/\xi)}$. 
If we wish to achieve $\varepsilon_{\rm LR}<\delta$ for a small number $\delta>0$, the parameter $l_0$ should be chosen as $l_0>\mathcal{O}(\xi\log{(1/\delta)})$. 
In other words, the compilation size $\tL$ should be taken as 
\begin{equation}
   \tL \gtrsim \mathcal{O}(\xi\log{(1/\delta)})+d_{H}+v\tau + 2d + 1. \label{eq:compile-size-2}
\end{equation}
This implies that the compilation size $\tL$ grows only logarithmically with the desired precision of $\varepsilon_{\rm LR}$.
Also, it should be noted that $\tL$ does not depend on $L$.

\section{Our proposal to calculate the Green's function}\label{sec:protocol}
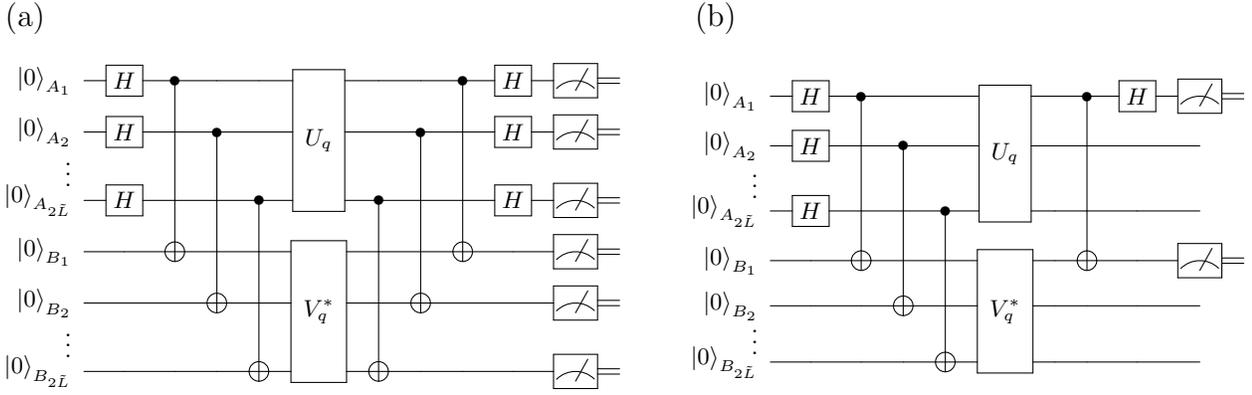
\begin{figure*}[t]
\leftline{{\large (a)} \hspace{0.47 \hsize} {\large (b)}}
\begin{minipage}{0.48\hsize}
\[
\Qcircuit @C=.9em @R=.7em {
  \lstick{\ket{0}_{A_1}} & \gate{H} & \ctrl{4} & \qw & \qw & \multigate{3}{U_q} & \qw & \qw & \ctrl{4} & \gate{H} & \meter & \cw \\
  \lstick{\ket{0}_{A_2}} & \gate{H} & \qw & \ctrl{4} & \qw & \ghost{U_q} & \qw & \ctrl{4} & \qw & \gate{H} & \meter & \cw \\
  \lstick{\vdots} & & & & & & & & & & & \\
  \lstick{\ket{0}_{A_{2\tL}}} & \gate{H} & \qw & \qw & \ctrl{4} & \ghost{U_q} & \ctrl{4} & \qw & \qw & \gate{H} & \meter & \cw \\
  \lstick{\ket{0}_{B_1}} & \qw & \targ  & \qw & \qw & \multigate{3}{V_q^*} & \qw & \qw & \targ  & \qw & \meter & \cw \\
  \lstick{\ket{0}_{B_2}} & \qw & \qw & \targ  & \qw & \ghost{V_q^*} & \qw & \targ  & \qw & \qw & \meter & \cw \\
  \lstick{\vdots} & & & & & & & & & &  \\
  \lstick{\ket{0}_{B_{2\tL}}} & \qw & \qw & \qw & \targ  & \ghost{V_q^*} & \targ  & \qw & \qw & \qw & \meter & \cw  \\
}
\]
\end{minipage}
\begin{minipage}{0.48\hsize}
\[
\Qcircuit @C=.9em @R=.7em {
  \lstick{\ket{0}_{A_1}} & \gate{H} & \ctrl{4} & \qw & \qw & \multigate{3}{U_q} & \qw & \ctrl{4} & \gate{H} & \meter & \cw \\
  \lstick{\ket{0}_{A_2}} & \gate{H} & \qw & \ctrl{4} & \qw & \ghost{U_q} & \qw & \qw & \qw & \qw \\
  \lstick{\vdots} & & & & & & & & & & \\
  \lstick{\ket{0}_{A_{2\tL}}} & \gate{H} & \qw & \qw & \ctrl{4} & \ghost{U_q} & \qw & \qw & \qw & \qw \\
  \lstick{\ket{0}_{B_1}} & \qw & \targ  & \qw & \qw & \multigate{3}{V_q^*} & \qw & \targ  & \qw & \meter & \cw \\
  \lstick{\ket{0}_{B_2}} & \qw & \qw & \targ  & \qw & \ghost{V_q^*} & \qw & \qw & \qw & \qw \\
  \lstick{\vdots} & & & & & & & & & &  \\
  \lstick{\ket{0}_{B_{2\tL}}} & \qw & \qw & \qw &  \targ  & \ghost{V_q^*} & \qw & \qw & \qw & \qw
}
\]
\end{minipage}
\caption{(a) Quantum circuit for measuring the cost function $C^f_{\mr{HST}}(U_f^{(\tL)}(\tau), V_f^{(\tL)}(\vec{\theta}))$ for the local $\tL$-sites fermionic system mapped into $2\tL$ qubits.
The probability to obtain the measurement outcome of all 0's for $2\tL$ qubits is equal to $1-C^f_{\mr{HST}}(U_f^{(\tL)}(\tau), V_f^{(\tL)}(\vec{\theta}))$.
(b) Quantum circuit for measuring the cost function $C_{\mr{LHST}}(U_f^{(\tL)}(\tau), V_f^{(\tL)}(\vec{\theta}))$ when using Jordan-Wigner transformation.
In this circuit, only the qubits $A_1$ and $B_1$ are measured, and the probability to find both qubits in the state $\ket{0}$ is equal to $1 - C_{\rm LHST}^{(1)}(U_f^{(\tL)}(\tau), V_f^{(\tL)}(\vec{\theta}))$. }
\label{fig:circuit-hst}
\end{figure*}

Here we describe a concrete protocol how to use LVQC to calculate the Green's function of (translationally-invariant) fermionic systems.
When simulating the total system of $L$ sites, $H^{(L)}_{f,\mr{PBC}}$, with the ansatz $V_f^{(L)}(\vec{\theta})$, our quantum-classical hybrid algorithm proceeds as follows.
\begin{enumerate}
 \item Define a local subsystem consisting of $\tL$ sites and specify the local Hamiltonian $H^{(\tL)}_{f,\mr{PBC}}$, its time evolution operator $U_{f}^{(\tL)}(\tau)$ with $\tau$ being a fixed time, and the local ansatz $V_f^{(\tL)}(\vec{\theta})$.
 \item Optimize the parameters $\vec{\theta}$ to minimize the cost functions $C^f_\mr{LHST}(U_{f}^{(\tL)}(\tau), V_f^{(\tL)}(\vec{\theta}))$ or $C^f_\mr{HST}(U_{f}^{(\tL)}(\tau), V_f^{(\tL)}(\vec{\theta}))$ by using the local $\tL$-sites system. 
 \item Utilize the ansatz of $L$ sites with the optimized $\vec{\theta}_\mr{opt}$, i.e., $V_f^{(L)}(\vec{\theta}_\mr{opt})$, in the calculation of the Green's function.
 More concretely, we approximate the time evolution operator $e^{-iHt}$ of the $L$-site system in Eq.~\eqref{eq:gf-pauli-kernel} (and Fig.~\ref{fig:circuit-gf}) at $t=n\tau \, (n=0,1,2,\cdots)$ as $e^{-i H (n\tau)} \approx (V_f^{(L)}(\vec{\theta}_\mr{opt}))^n$ and calculate the Green's function by using the $L$-sites system. 
\end{enumerate}
We have several remarks for these steps.

First, in the step 1, the compilation size $\tL$ and the fixed time $\tau$ are set to sufficiently suppress the error $\varepsilon_{\mr{LR}}$ from the LR bound in Eqs.~\eqref{eq:cost-lhst-lvqc} and \eqref{eq:cost-hst-lvqc}.
Although we do not know exact values of the several numbers in the LR bound such as the velocity $v$ or the length $\xi$ \textit{a priori},
one can choose $\tL$ and $\tau$ just by hand or by performing the benchmark calculation for the smaller system $l < L$ in practice.  
Moreover, since the LVQC theorem gives only an upper bound of the error, it may be possible to choose $\tL$ which does not satisfy Eq.~\eqref{eq:compile-size} and still have the small $\varepsilon_{\mr{LR}}$ in actual systems, as we will numerically see in the next section.
We note that at least we can suppress the error $\varepsilon_{\mr{LR}}$ exponentially by increasing $l_0$ or $\tL$.

Second, in the step 2, the cost functions $C_\mr{LHST}^f(U_{f}^{(\tL)}(\tau), V_f^{(\tL)}(\vec{\theta}))$ or $C_\mr{HST}^f(U_{f}^{(\tL)}(\tau), V_f^{(\tL)}(\vec{\theta}))$ in the local $\tL$-sites system must be computed during the optimization.
When using a quantum computer made of qubits, we map the local system into $2\times 2\tL = 4\tL$ qubits and obtain the qubit representations of $U_{f}^{(\tL)}(\tau)$ and $V_f^{(\tL)}(\vec{\theta})$, denoted as $U_q$ and $V_q$ respectively.
The two cost functions can be evaluated by quantum circuits depicted in Fig.~\ref{fig:circuit-hst}
(when we use mapping other than the Jordan-Wigner transformation, the circuits must be slightly modified).
We also add that we can use the mixed cost function like
\begin{equation}
  C_{\alpha}^f(U_f, V_f) = \alpha C^f_\mr{HST}(U_f,V_f) + (1-\alpha) C^f_\mr{LHST}(U_f,V_f)
\end{equation}
with some $0 \leq \alpha \leq 1$ to alleviate the so-called barren plateau problem~\cite{Khatri-QAQC}.

Third, in step 3, we repeatedly apply the approximation of the time evolution operator for $\tau$, $(V^{(L)}(\vec{\theta}))^n \approx (e^{-iH\tau})^n = e^{-iH (n\tau)}$.
The maximum number of $n=0,1,2, \cdots$ depends on the time scale at which we want to simulate and the noise level of quantum hardwares to execute the quantum circuit in Fig.~\ref{fig:circuit-gf}.
We provide a concrete estimate of the maximum number of $n$ for the fermion Hubbard model in Sec.~\ref{sec:ResourceEstimation}.

\section{Numerical demonstration for Fermi-Hubbard model}\label{sec:Numerical}
\begin{figure}[tbp]
    \includegraphics[keepaspectratio, scale=0.44]{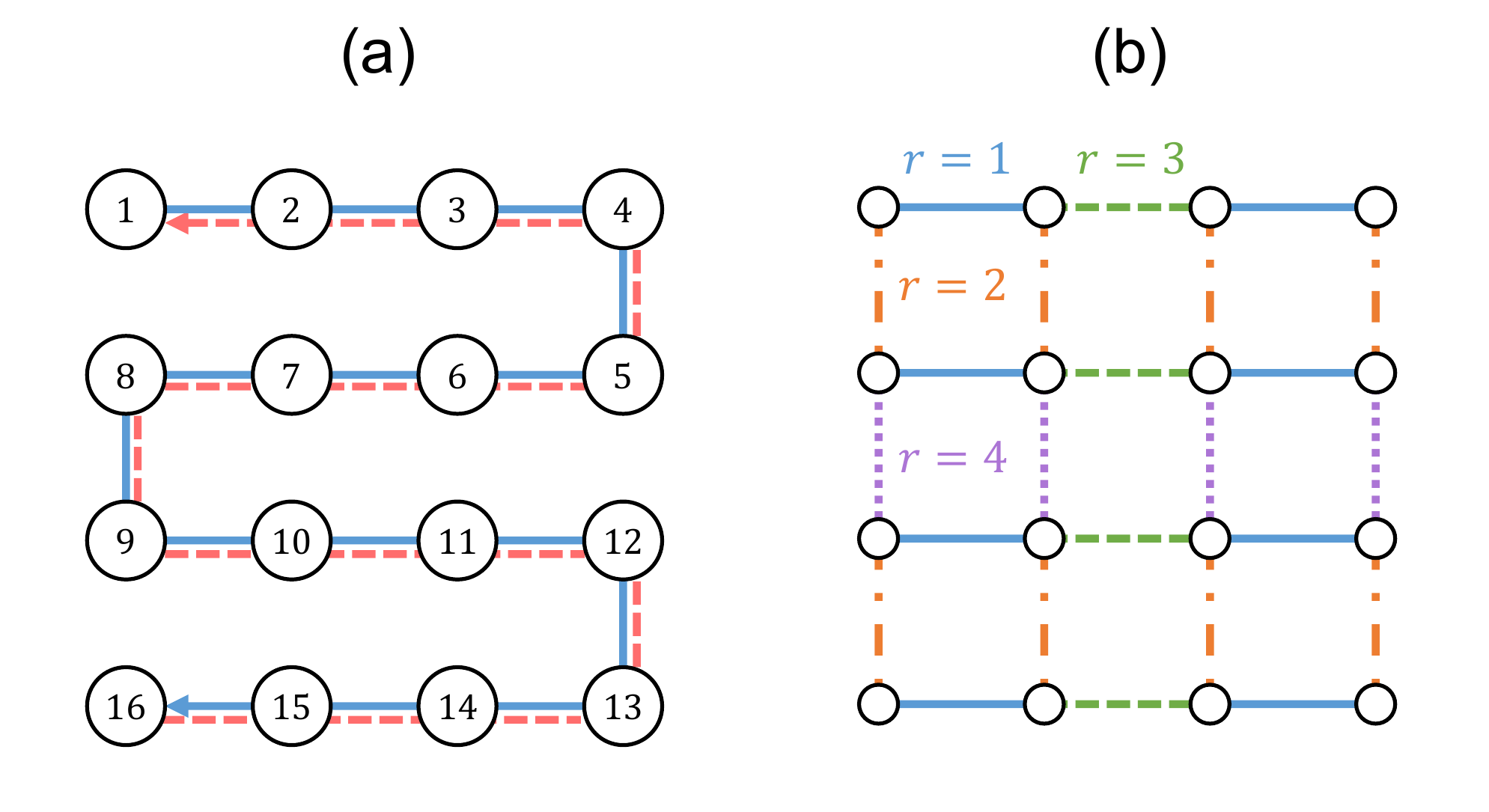}
    \caption{(a) Illustration of how fermionic modes are mapped to qubits under the Jordan-Wigner transformation in a $4\times 4$ lattice Fermi-Hubbard model. The numbers in the circles denote the site index $i$. The blue solid (red dashed) line represents the order of Jordan-Wigner encoding for spin-up (spin-down) fermions. (b) Illustration of the four sets of the hopping terms. The blue solid (green dashed) line represents the horizontal hopping terms with $r=1$ ($r=3$), while the orange dash-dotted (purple dotted) line represents the vertical hopping term with $r=2$ ($r=4$).} 
    \label{fig:jordan-wigner}
\end{figure}

In this section, we perform a numerical simulation of the Green's function using the LVQC algorithm. 
We consider the Fermi-Hubbard model on a $L_x\times L_y$ lattice under periodic boundary condition 
\begin{align}
    H^{(L)} &= -t\sum_{\substack{\langle i,j \rangle\\ 1\leq i,j\leq L}}\sum_{\sigma=\uparrow,\downarrow}(c_{i\sigma}^{\dag}c_{j\sigma}+\mathrm{H.c.}) \nonumber\\
  &
  + U\sum_{i=1}^{L}n_{i\uparrow}n_{i\downarrow} -\mu\sum_{i=1}^{L}\sum_{\sigma=\uparrow,\downarrow}n_{i\sigma},
  \label{eq:fermi-hubbard}
\end{align}
where $c_{i\sigma}$ ($c_{i\sigma}^{\dag}$) is the creation (annihilation) operator of a fermion at site $i(=1,2,\cdots,L)$ with spin $\sigma(=\uparrow,\downarrow)$ on a lattice having $L(=L_xL_y)$ sites, $\langle i,j \rangle$ denotes neighboring sites, and $n_{i\sigma}=c_{i\sigma}^{\dag}c_{i\sigma}$ is the fermion density operator. 
The parameters $t$, $U$, and $\mu$ represent hopping integral, on-site Coulomb interaction, and chemical potential, respectively. 
We set $t=1$ throughout this paper. 
Note that the Fermi-Hubbard Hamiltonian~\eqref{eq:fermi-hubbard} is local and exhibits the LR bound~\cite{PRXQuantum.1.010303}, and hence the LVQC algorithm is applicable. 
We map the fermionic Hamiltonian~\eqref{eq:fermi-hubbard} to the qubit one by adopting the Jordan-Wigner transformation~\cite{Jordan1928}, 
\begin{align}
 \begin{split}
     c_{i\sigma} &\mapsto \frac{1}{2}(X_{i_{\sigma}}+iY_{i_{\sigma}})Z_{i_\sigma-1}^{\to}, \\
    c_{i\sigma}^{\dag} &\mapsto \frac{1}{2}(X_{i_{\sigma}}-iY_{i_{\sigma}})Z_{i_\sigma-1}^{\to}, 
 \end{split} \label{eq:jordan-wigner}
\end{align}
where $X_{i_\sigma}$, $Y_{i_\sigma}$, and $Z_{i_\sigma}$ are the Pauli matrices at qubit $i_\sigma$, and $Z_{i-1}^{\to}\equiv\prod_{k<i}Z_k$. 
We use the so-called snaked-shaped configuration where the qubits are ordered as $i_{\uparrow}=i(=1,2,\cdots,L)$ and $i_{\downarrow}=2L+1-i(=2L,2L-1,\cdots,L+1)$ as shown in Fig.~\ref{fig:jordan-wigner}(a). 
The total number of qubits is $N=2L$. 
Note that the form of Eq.~\eqref{eq:jordan-wigner} is consistent with Eq.~\eqref{eq:qubit-map}. 
Then, the Fermi-Hubbard Hamiltonian in the qubit representation is described as
\begin{align}
    H^{(L)} &= -\frac{t}{2}\sum_{\substack{\langle i,j \rangle\\ 1\leq i,j\leq L}}\sum_{\sigma=\uparrow,\downarrow}
    (X_{i_\sigma}X_{j_\sigma}+Y_{i_\sigma}Y_{j_\sigma})Z_{i_\sigma,j_\sigma}^{\leftrightarrow} \nonumber\\
    & + \frac{U}{4}\sum_{i=1}^{L}Z_{i_\uparrow}Z_{i_\downarrow} + \frac{1}{2}\left(\mu-\frac{U}{2}\right)\sum_{i=1}^{L}\sum_{\sigma=\uparrow,\downarrow}Z_{i_\sigma},
    \label{eq:fermi-hubbard-qubit}
\end{align}
where $Z_{i,j}^{\leftrightarrow}$ is the Jordan-Wigner string defined as 
\begin{equation}
    Z_{i,j}^{\leftrightarrow} = 
    \begin{cases}
        1 & (i = j \pm 1 ) \\
        \prod_{i< k < j}Z_{k} & (i < j-1) \\
        \prod_{i> k > j}Z_{k} & (i> j+1)
    \end{cases}.
\end{equation}
To calculate the Green's function of the Fermi-Hubbard model by using the method described in Sec.~\ref{sec:Green} and \ref{sec:LVQC}, we first prepare the ground state of the model~\eqref{eq:fermi-hubbard}.
In this paper, we prepare the ground state by the exact diagonalization based on the Lanczos algorithm. 
Then, we optimize the HST or LHST cost functions with the so-called variational Hamiltonian ansatz~\cite{Wecker2015,reiner2019finding} inspired by the Trotter decomposition of the time-evolution operator.
To efficiently construct the variational Hamiltonian ansatz, we split the model~\eqref{eq:fermi-hubbard-qubit} into parts that consist of terms that are sums of commuting components~\cite{Cade-hubbardVQE-2020}
\begin{equation}
    H^{(L)} = -\frac{t}{2}\sum_{r=1}^{4}P_{t_r}^{(L)} 
    +\frac{U}{4}P_{U}^{(L)} +\frac{\mu}{2}P_{\mu}^{(L)} ,
\end{equation}
where
\begin{align}
    P_{t_r}^{(L)}  &= \sum_{\sigma}\sum_{\substack{\langle i,j\rangle_r\\ 1 \leq i,j \leq L}}(X_{i_\sigma}X_{j_\sigma}+Y_{i_\sigma}Y_{j_\sigma})Z_{i_\sigma,j_\sigma}^{\leftrightarrow},\\
    P_{U}^{(L)}  &= \sum_{i=1}^{L}(Z_{i_\uparrow}Z_{i_\downarrow}-Z_{i_\uparrow}-Z_{i_\downarrow}),\\
    P_{\mu}^{(L)}  &= \sum_{i=1}^{L}\sum_{\sigma=\uparrow,\downarrow}Z_{i_\sigma} ,
\end{align}
and $\langle i,j\rangle_r$ represents the horizontal $(r=1,3)$  and vertical $(r=2,4)$ hopping network as shown in Fig.~\ref{fig:jordan-wigner}(b).
Note that the horizontal hopping term $P_{t_{1,3}}^{(L)}$ (the vertical hopping terms $P_{t_{2,4}}^{(L)}$) vanishes on a one-dimensional lattice with $L_x=1$ ($L_y=1$). 
Then, we construct the variational Hamiltonian ansatz as
\begin{equation}
    V^{(L)}_{d}(\vec{\theta}) = \prod_{k=1}^{d}\left[ 
    e^{i\theta_1^{(k)} P_{\mu}^{(L)}}
    e^{i\theta_2^{(k)} P_{U}^{(L)}}
    \prod_{r=1}^{4}e^{i\theta_3^{(k)} P_{t_r}^{(L)}}
    \right],
    \label{eq:ansatz-fermi-hubbard}
\end{equation}
where $d$ denotes the depth of the ansatz and $\vec{\theta}=\{\theta_{j=1,2,3}^{(k)}\}$ denotes the variational parameters. 
Considering the translation symmetry of the model, we set all of the hopping terms $P_{t_r}^{(L)}$ to have uniform variational parameters $\theta_{3}^{(k)}$. 
Thus, the total number of the variational parameters of the ansatz~\eqref{eq:ansatz-fermi-hubbard} is $3d$. 
Note that the ansatz~\eqref{eq:ansatz-fermi-hubbard} respects symmetries and locality of the original fermionic Hamiltonian~\eqref{eq:fermi-hubbard}, and hence the LVQC algorithm is applicable (see Appendix~\ref{append:LVQC-fermion}).

\begin{figure}[tbp]
    \includegraphics[keepaspectratio, scale=0.38]{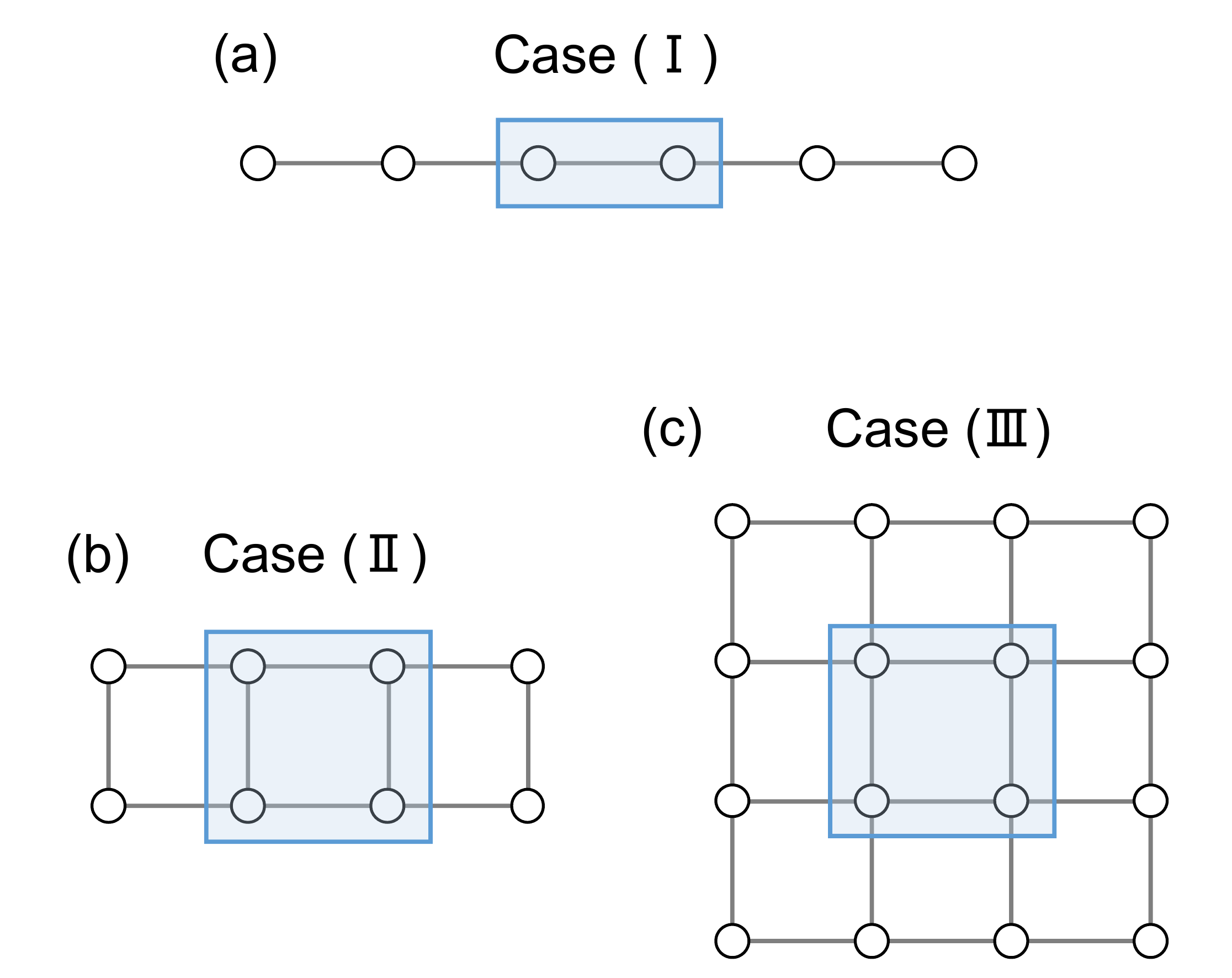}
    \caption{An illustration of the relation of the whole lattice size $L$ and the compilation size $\tilde{L}$. The blue-colored area represents the compilation size. 
    (a) $\tilde{L}=2\times1$ in a $L=6\times1$ lattice. 
    (b) $\tilde{L}=2\times2$ in a $L=4\times2$ lattice. 
    (c) $\tilde{L}=2\times2$ in a $L=4\times4$ lattice. } 
    \label{fig:compile-size}
\end{figure}

Under the above setup, we perform numerical simulations by classical computers with the fast quantum circuit library Qulacs~\cite{qulacs}.  
We compute the cost function $C_{\rm LHST}^f$ for the $\tilde{L}_x\times\tilde{L}_y$ lattice Fermi-Hubbard model based on the circuits shown in Fig.~\ref{fig:circuit-hst}, where the lattice size of the compilation is denoted as $\tilde{L}=\tilde{L}_x\times\tilde{L}_y$. 
Note that we choose to train only $C_{\rm LHST}^f$ and not $C_{\rm HST}^f$ because $C_{\rm HST}^f$ is suggested to have an apparent barren plateau issue~\cite{Khatri-QAQC}. 
Examples of taking the compilation size $\tilde{L}$ are illustrated in Fig.~\ref{fig:compile-size}. 
In Sec.~\ref{subsec:hubbard-1d} and~\ref{subsec:hubbard-2d}, we show the numerical results of Green's function and spectral function obtained by using the LVQC algorithm with the three patterns of compilation size shown in Fig.~\ref{fig:compile-size}. 
To implement the target unitary $e^{-iH^{(\tilde{L})}\tau}$, 
we utilize the (first-order) Trotter decomposition 
\begin{equation}
    U_{d}^{(\tilde{L})}(\tau) = \left(
    e^{-i\frac{\mu}{2}P_{\mu}^{(\tilde{L})}\frac{\tau}{d}}
    e^{-i\frac{U}{4}P_{U}^{(\tilde{L})}\frac{\tau}{d}} \prod_{r=1}^{4}e^{i\frac{t}{2}P_{t_r}^{(\tilde{L})}\frac{\tau}{d}}
    \right)^{d},
    \label{eq:trotter}
\end{equation}
with a sufficiently large depth $d=100$. 
The cost functions are minimized by using the Broyden-Fletcher-Goldfarb-Shanno (BFGS) method implemented in SciPy~\cite{SciPy} with the maximum iteration set to 128. 
The initial parameter set $\vec{\theta}_0$ is chosen so as to the initial ansatz $V^{(\tilde{L})}_{d}(\vec{\theta}_0)$ becomes equivalent to the Trotter decomposition with the same depth $d$, i.e., $V^{(\tilde{L})}_{d}(\vec{\theta}_0)\equiv U_{d}^{(\tilde{L})}(\tau)$. 
Using the optimized parameter set $\vec{\theta}_{\rm opt}$ obtained by the above local compilation procedure, the real-space Green's function $G_{i\sigma,i'\sigma'}^{\rm R}(t)$ is calculated based on Eq.~\eqref{eq:gf-pauli}.  
The momentum space representation of the Green's function $G^{\rm R}_{\mathbf{k}\sigma}(t)$ is calculated based on Eq.~\eqref{eq:gf-momentum} replacing the spatial coordinate $\mathbf{x}$ with the site index $i$. 
Then, the spectral function is obtained using Eq.~\eqref{eq:green-fourier}. 
To numerically evaluate the integral in Eq.~\eqref{eq:green-fourier}, we first discretize the time domain as $t=\{0,\Delta{t},2\Delta{t},\cdots,N_t\Delta{t}\}$ with a small time step $\Delta{t}$ and a large integer $N_t$. 
Then, approximate the integral in Eq.~\eqref{eq:green-fourier} by using the trapezoidal rule, 
\begin{align}
    \tilde{G}^{\rm R}_{\mathbf{k}\sigma}(\omega)
    &\approx \frac{\Delta{t}}{2}\left(f_0 + 2\sum_{\ell=1}^{N_t-1} f_{\ell} + f_{N_{t}} \right),
    \label{eq:green-fourier-sum}
\end{align}
where
\begin{equation}
    f_{\ell} \equiv e^{i(\omega+i\eta)\ell\Delta{t}}G^{\rm R}_{\mathbf{k}\sigma}(\ell\Delta{t}) .
\end{equation}
In the following numerical calculations, we set $\mu=U/2$ and consider the half-filling regime where the particle-hole symmetry is preserved.

\subsection{One-dimensional case}\label{subsec:hubbard-1d}
\begin{figure}[tbp]
    \includegraphics[keepaspectratio, scale=0.63]{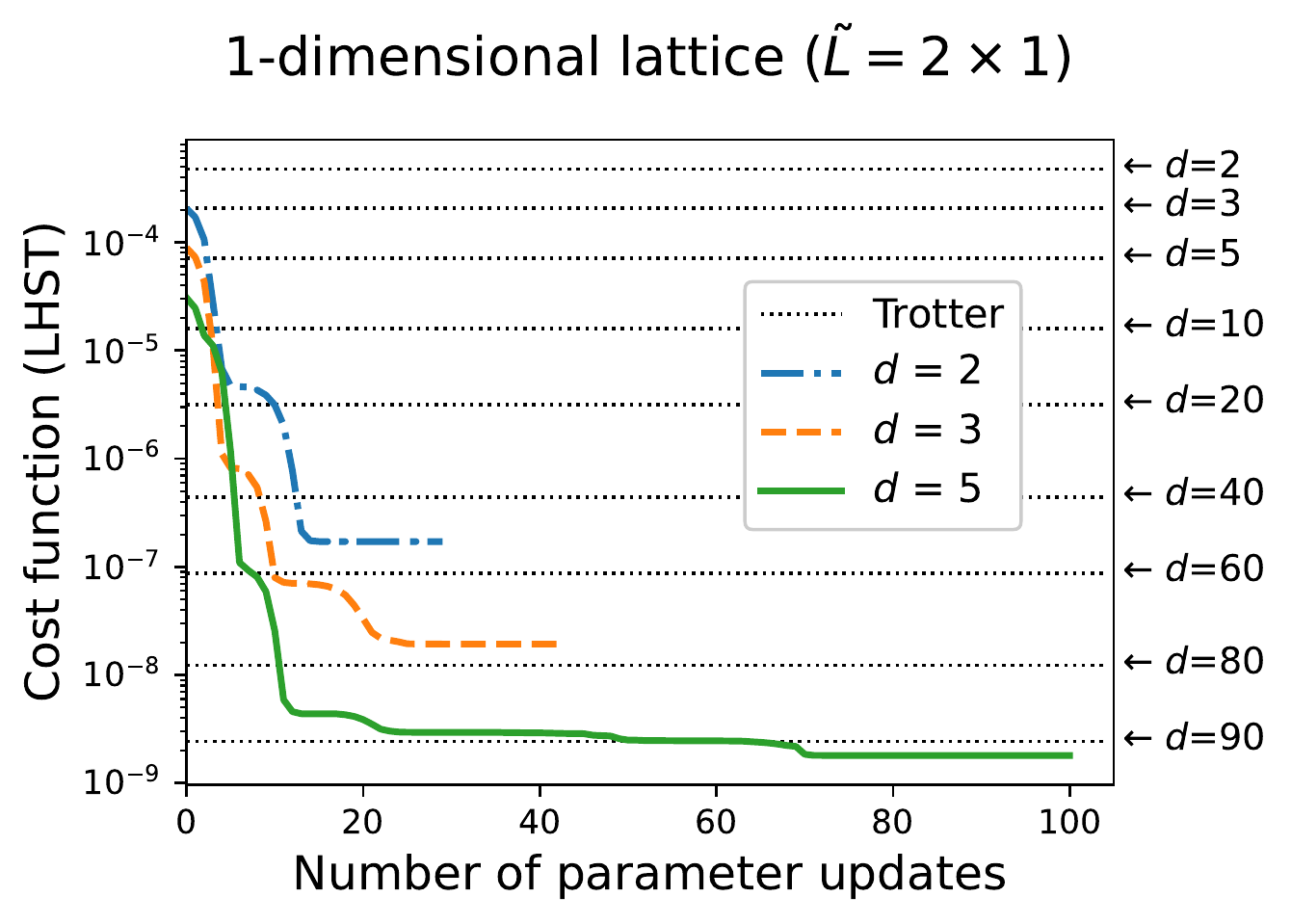}
    \caption{
    The history of optimization of the cost function $C_{\rm LHST}^f(U_{\rm 100}^{(\tilde{L})}(\tau),V_{d}^{(\tilde{L})}(\vec{\theta}))$ at the compilation size $\tilde{L}=2\times1$. 
    The parameters are set to be $U=10$ and $\tau=0.1$. The blue dash-dotted line, orange dashed line, and green solid line represent the results for $d=2,3,5$, respectively. The black dotted lines represent the corresponding cost function for the Trotter decomposition $C_{\rm LHST}^f(U_{100}^{(\tilde{L})}(\tau),V_{d}^{(\tilde{L})}(\vec{\theta}_0))$ with various values of the depth $d$. } 
    \label{fig:1d-cost}
\end{figure}

First, we show the numerical result for Case (I) shown in Fig.~\ref{fig:compile-size}(a).
Figure~\ref{fig:1d-cost} shows the history of the cost function $C_{\rm LHST}^f(U_{\rm 100}^{(\tilde{L})}(\tau),V_{d}^{(\tilde{L})}(\vec{\theta}))$ during the optimization at the compilation size $\tilde{L}=2\times 1$ and time $\tau=0.1$ for $d=2$ (blue dash-dotted line), $d=3$ (orange dashed line), and $d=5$ (green solid line). 
For comparison, we show the value of the cost functions for the Trotter decomposition, i.e., $C_{\rm LHST}^f(U_{100}^{(\tilde{L})}(\tau),V_{d}^{(\tilde{L})}(\vec{\theta}_0))=C_{\rm LHST}^f(U_{\rm 100}^{(\tilde{L})}(\tau),U_{d}^{(\tilde{L})}(\tau))$, for various values of the depth $d$ (black dotted lines). 
For each depth, $C_{\rm LHST}^f(U_{\rm 100}^{(\tilde{L})}(\tau),V_{d}^{(\tilde{L})}(\vec{\theta}))$ converges to the optimal value, which is significantly smaller than that of the corresponding Trotter decomposition. 
For example, the optimized value of the cost function $C_{\rm LHST}^f(U_{\rm 100}^{(\tilde{L})}(\tau),V_{d}^{(\tilde{L})}(\vec{\theta}_{\rm opt}))$ for $d=5$ is $1.80\times10^{-9}$, which is smaller than that of the depth-80 Trotter decomposition $5.31\times10^{-9}$. 
This can also be verified in terms of the average gate fidelity, 
\begin{align}
    \bar{F}(U,V) &= \int_{\psi}d\psi |\bra{\psi}V^{\dag}U\ket{\psi}|^2, \label{eq:fidelity-def} \\
    &= \frac{1}{D+1}\left(1+\frac{1}{D}|\mathrm{Tr}(V^{\dag}U)|^2 \right),
    \label{eq:fidelity-comp}
\end{align}
where $U$ and $V$ are assumed to act on a $D$-dimensional space, and the integral in Eq.~\eqref{eq:fidelity-def} is taken over all states $\ket{\psi}$ chosen according to the Haar measure. 
Using Eq.~\eqref{eq:fidelity-comp}, we obtain that the optimized ansatz with $d=5$ has $\bar{F}(U_{\rm 100}^{(\tilde{L})}(\tau),V_{d}^{(\tilde{L})}(\vec{\theta}_{\rm opt})) = 1-4.83\times10^{-9}$, which is comparable to the average gate fidelity of the depth-80 Trotter circuit $\bar{F}(U_{\rm 100}^{(\tilde{L})}(\tau),V_{80}^{(\tilde{L})}(\vec{\theta}_{0}))=1-1.31\times10^{-9}$. 
These results indicate that the number of gates needed to accurately approximate the time evolution operator at the compilation size $\tilde{L}=2\times1$ is successfully reduced to less than $5/80=1/16$ compared to the Trotter decomposition. 

\begin{figure*}[tbp]
    \includegraphics[keepaspectratio, scale=0.39]{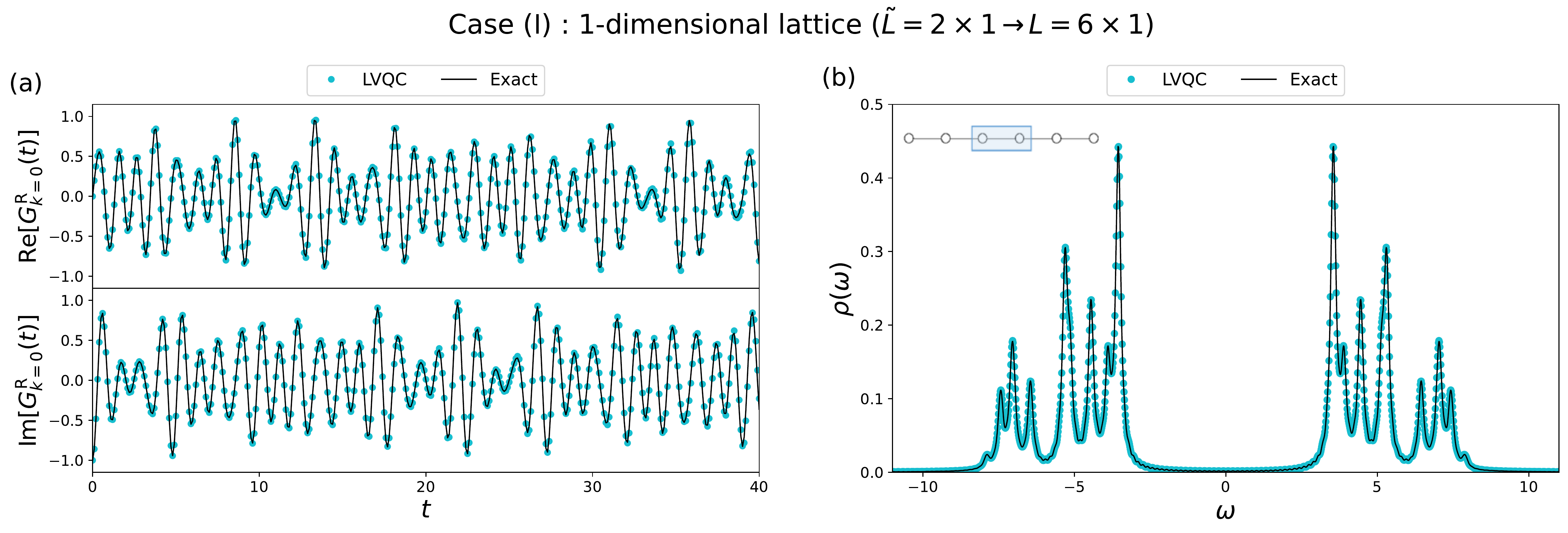}
    \caption{(a) The Green's function $G_{\mathbf{k}}^{\rm R}(t)$ at $\mathbf{k}=0$ and (b) the DOS $\rho(\omega)$ for the one-dimensional Fermi-Hubbard model on a $6\times 1$ lattice at $U=10$ and half-filling. 
    The black lines and cyan dots represent the results of the exact diagonalization and the LVQC algorithm, respectively.
    The LVQC algorithm is performed by optimizing the local cost function $C_{\rm LHST}^f$ with the compilation size $\tilde{L}=2\times 1$, time $\tau=0.1$, and depth $d=5$. 
    The Green's function is calculated in the time domain $t\in [0,40]$ with the step $\tau=0.1$. We take $\eta=0.1$ for the calculation of the DOS. } 
    \label{fig:lvqc-1d}
\end{figure*}

Next, we examine the Green's function calculated using the LVQC algorithm. 
Figure~\ref{fig:lvqc-1d} shows the Green's function of the $6\times1$ lattice Fermi-Hubbard model obtained by the LVQC (cyan dots) and exact diagonalization (black solid line).  
The LVQC result is calculated by using the optimal parameter set $\vec{\theta}_{\rm opt}$ obtained by minimizing the local cost function $C_{\rm LHST}^f(U_{\rm 100}^{(\tilde{L})}(\tau),V_{d}^{(\tilde{L})}(\vec{\theta}))$ at the compilation size $\tilde{L}=2\times1$ and time $\tau=0.1$ with the ansatz depth $d=5$. 
Note that the average gate fidelity of the optimized circuit $V^{(L)}_{d}(\vec{\theta}_{\rm opt})$ is still better than the corresponding Trotter circuit even when the optimized ansatz extended to the whole lattice size $L=6\times1$.
Indeed, the average gate fidelity of the LVQC circuit is $\bar{F}(U_{\rm 100}^{(L)}(\tau),V_{d}^{(L)}(\vec{\theta}_{\rm opt})) = 1-6.62\times10^{-6}$ for $d=5$, which is comparable to that of the depth-30 Trotter circuit $\bar{F}(U_{\rm 100}^{(L)}(\tau),V_{30}^{(L)}(\vec{\theta}_{0}))=1-3.75\times10^{-6}$.
The long-time-scale dynamics at the time $t=n\tau$ ($n \in \mathbb{N}$) is calculated by repeatedly applying the optimized circuit $V^{(L)}_{d}(\vec{\theta}_{\rm opt})$ as $(V_{d}^{(L)}(\vec{\theta}_{\rm opt}))^n\approx e^{-iH^{(L)}n\tau}$. 
As shown in Fig.~\ref{fig:lvqc-1d}(a), the LVQC algorithm nicely reproduces the exact Green's function for a long time scale up to $t=400\tau$. 
Consequently, the DOS, which is obtained by the Fourier transform of the Green's function in the time domain (see Eq.~\eqref{eq:dos}), is also well reproduced by the LVQC method as shown in Fig.~\ref{fig:lvqc-1d}(b). 

Here, we investigate the accuracy of the LVQC algorithm for computing the Green's function. 
We first examine the dependence of the accuracy on the depth of the ansatz $d$, which is essential to increase the expressive power of the ansatz and reduce optimization errors $\varepsilon_{\rm LHST}$ and $\varepsilon_{\rm HST}$. 
To see the accuracy of the LVQC algorithm, we examine the absolute error (AE) of the Green's function 
\begin{equation}
    \delta G_{\mathbf{k}}(t) = \left| G^{\rm R,exact}_{\mathbf{k}}(t) - G^{\rm R,approx}_{\mathbf{k}}(t) \right|, 
    \label{eq:error-gf}
\end{equation}
where $G^{\rm R,exact}_{\mathbf{k}}$ and $G^{\rm R,approx}_{\mathbf{k}}$ are the exact and approximate value of the Green's function, respectively. 
Because of the symmetry of the up and down spins in the Fermi-Hubbard model, we consider only the up spin component of the Green's function, and spin index $\sigma$ is omitted, i.e., $G^{\rm R}_{\mathbf{k}}(t)\equiv G^{\rm R}_{\mathbf{k}\uparrow}(t)$.
\begin{figure*}[tbp]
    \includegraphics[keepaspectratio, scale=0.39]{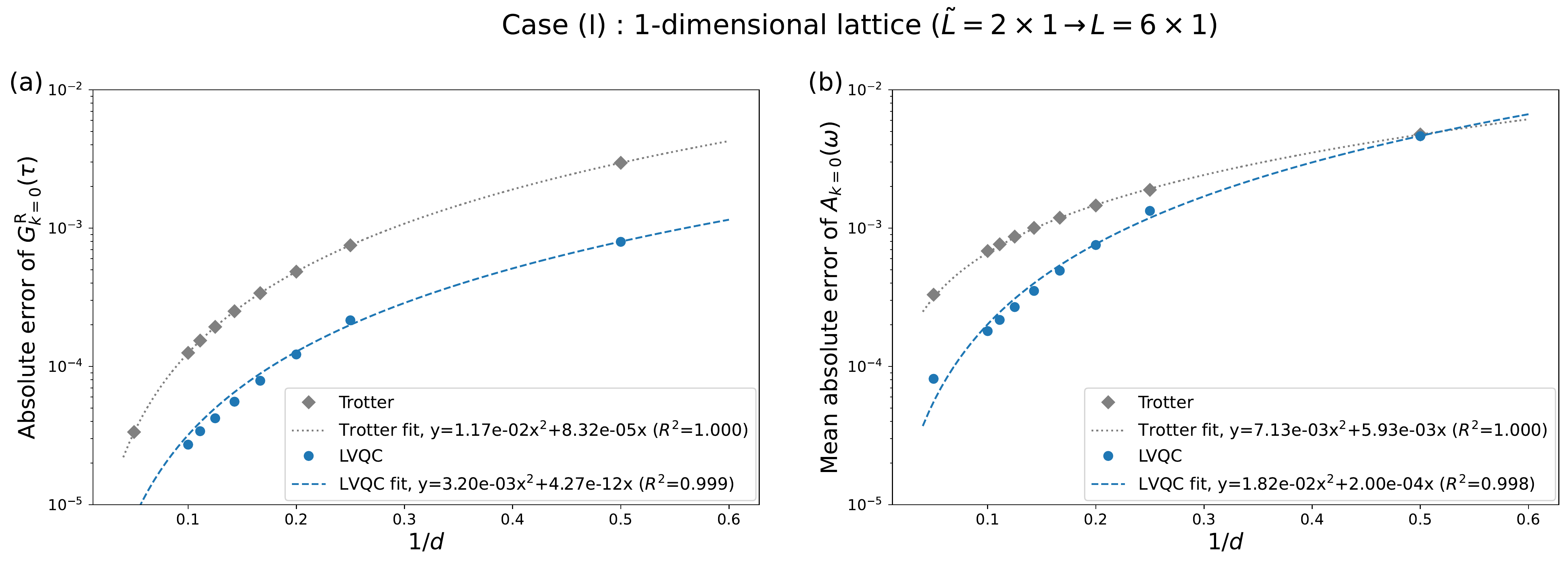}
    \caption{(a) The AE of the Green's function $\delta G_{\mathbf{k}=0}(\tau=0.1)$ and (b) the MAE of the spectral function $\delta A_{\mathbf{k}=0}$ for the one-dimensional Fermi-Hubbard model on a $6\times 1$ lattice at $U=10$ and half-filling. 
    The spectral function is calculated by the Fourier transformation of the Green's function $G_{\mathbf{k}}^{\rm R}(t)$ at $t\in[0,50]$ with the step $\tau=0.1$. 
    The broadening is taken as $\eta=0.1$. The MAE is calculated by setting $\omega_{\rm c}=15$ and $N_\omega=1000$. 
    The blue dots represent the results obtained by the LVQC method with the compilation size $\tilde{L} = 2\times 1$ and time $\tau=0.1$. The optimization is performed for the local cost function $C_{\rm LHST}^f$.
    The grey diamonds represent the results obtained by the Trotter decomposition of the time evolution operator for the $6\times 1$ lattice. 
    The blue dashed (grey dotted) line represents the fitting of the data of the LVQC (Trotter decomposition) by a quadratic function $y=\alpha x^2+\beta x$ based on the non-linear least squares. 
    The values of the coefficients $\alpha$, $\beta$ and the coefficient of determination $R^2$ are shown in the legend. } 
    \label{fig:1d-depth}
\end{figure*}
Figure~\ref{fig:1d-depth}(a) shows the AE of the Green's function $\delta G_{\mathbf{k}}(t)$ at $t=\tau$ and $\mathbf{k}=0$ as a function of the inverse of the depth $d$. 
The exact value $G^{\rm R,exact}_{\mathbf{k}}$ is prepared by the exact diagonalization, while the approximate value $G^{\rm R,approx}_{\mathbf{k}}$ is computed by the LVQC method with $\vec{\theta} = \vec{\theta}_{\rm opt}$ (blue circles) or the corresponding Trotter decomposition with $\vec{\theta} = \vec{\theta}_0$ (grey diamonds). 
As seen from Fig.~\ref{fig:1d-depth}(a), the AE $\delta G_{\mathbf{k}=0}(\tau)$ of the LVQC method is much smaller than that of the Trotter decomposition in a wide range of $d$. 
For example, the value of $\delta G_{\mathbf{k}=0}(\tau)$ of the LVQC method with $d=5$ is $1.22\times10^{-4}$, which is about 4 times smaller than the corresponding value of the Trotter decomposition $4.84\times10^{-4}$. 
On the other hand, the AE $\delta G_{\mathbf{k}=0}(\tau)$ of the LVQC and Trotter decomposition exhibit similar $1/d$ dependence and monotonically decreases with increasing the depth $d$. 
The $1/d$ dependence for both LVQC and Trotter decomposition are well fitted in the form of $\alpha/d^2+\beta/d$, where $\alpha$ and $\beta$ are positive fitting parameters.
Although the $1/d$ dependence of the AE of the Trotter decomposition is different from the well-known scaling of $\mathcal{O}(1/d)$~\cite{Lloyd1996universal}, it is consistent with some previous studies reporting the Trotter error in the form of $\mathcal{O}(\alpha/d^2+\beta/d)$~\cite{Tran-trotter-2020,Layden-trotter-prl2022}. 
Therefore, we can interpret that the error of the LVQC method inherits the scaling property of the Trotter error owing to the similarity between the variational Hamiltonian ansatz and the Trotter circuit. 
Similar behavior is observed in the mean absolute error (MAE) of the spectral function in the region of $\omega \in [-\omega_{\rm c}, \omega_{\rm c}]$,
\begin{equation}
    \delta A_{\mathbf{k}} = \frac{1}{2N_\omega+1}\sum_{n=-N_\omega}^{N_\omega}\left| A^{\rm exact}_{\mathbf{k}}(\omega_n) - A^{\rm approx}_{\mathbf{k}}(\omega_n) \right|, 
    \label{eq:error-spectral}
\end{equation}
where $2N_\omega+1$ is the total number of data points, $\omega_n=\omega_{\rm c}n/N_{\omega}$, and $A^{\rm exact}_{\mathbf{k}}$ ($A^{\rm approx}_{\mathbf{k}}$) is the exact (approximate) value of the spectral function. 
Figure~\ref{fig:1d-depth}(b) shows the MAE of the spectral function $\delta A_{\mathbf{k}}$ at $\mathbf{k}=0$ as a function of the inverse of the depth $d$. 
We take $\omega_{\rm c}=15$ and $N_{\omega}=1000$. 
The exact value $A^{\rm exact}_{\mathbf{k}}$ is prepared by the exact diagonalization, while the approximate value $A^{\rm approx}_{\mathbf{k}}$ is computed by the LVQC method with $\vec{\theta} = \vec{\theta}_{\rm opt}$ (blue circles) or the corresponding Trotter decomposition with $\vec{\theta} = \vec{\theta}_0$ (grey diamonds). 
The MAE $\delta A_{\mathbf{k}=0}$ of the LVQC method is smaller than that of the Trotter decomposition in a wide range of $d$. 
For example, the value of $\delta A_{\mathbf{k}=0}$ of the LVQC method with $d=5$ is $7.55\times10^{-4}$, which is about two times smaller than the corresponding value of the Trotter decomposition $1.46\times10^{-3}$. 
Note that the accuracy of the LVQC method in terms of $\delta A_{\mathbf{k}}$ is not so good as that in terms of $\delta G_{\mathbf{k}}(\tau)$. 
This is due to the accumulation of errors in a large time regime by repeatedly applying the approximate time evolution operator. 
On the other hand, the $1/d$ dependence of the MAE $\delta A_{\mathbf{k}}$ for both LVQC and Trotter decomposition are well fitted in the form of $\alpha/d^2+\beta/d$ in the same way as the AE $\delta G_{\mathbf{k}}$. 

We also investigate the dependence of the accuracy on the system size by fixing the compilation size $\tilde{L}$. 
\begin{figure}[tbp]
    \includegraphics[keepaspectratio, scale=0.4]{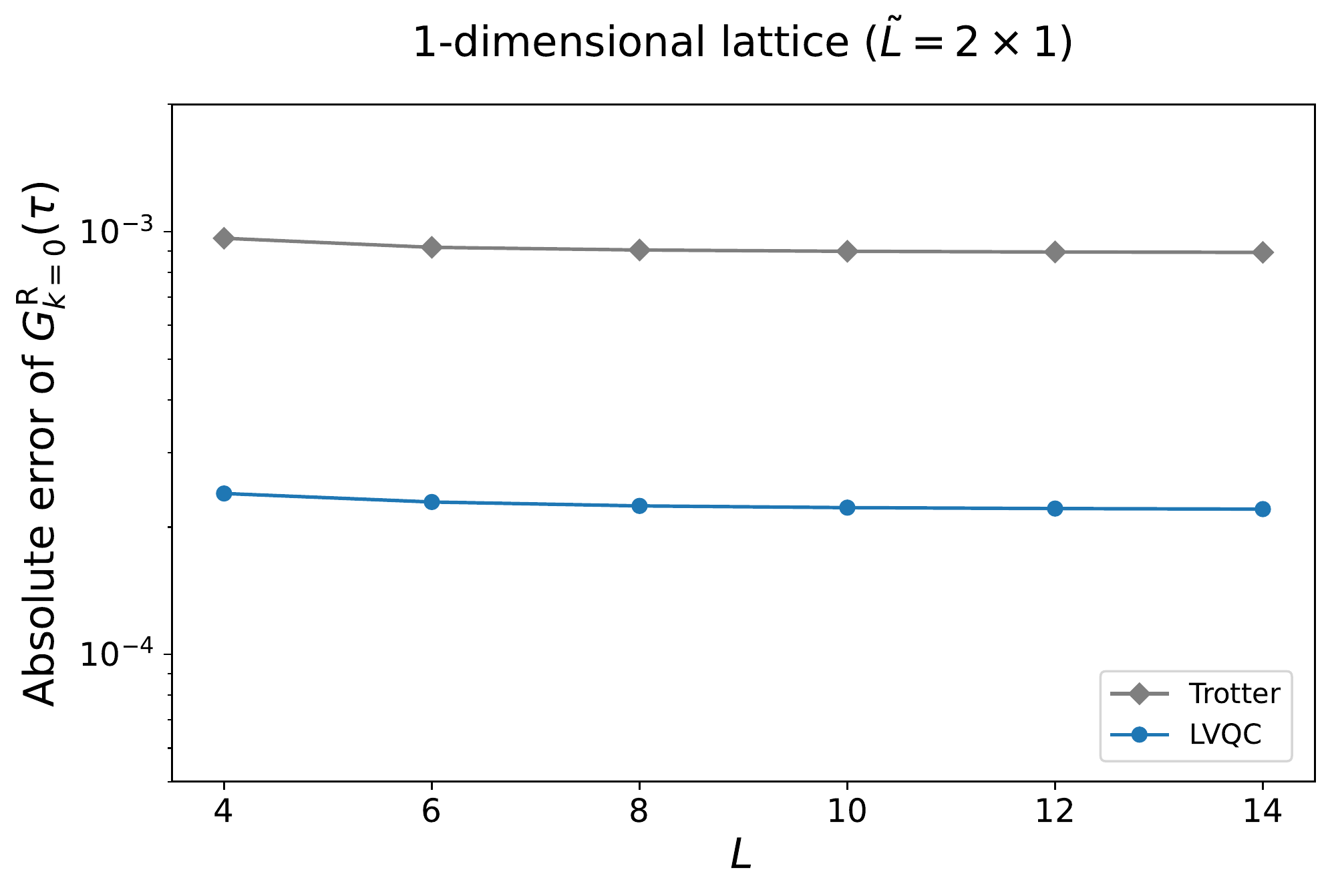}
    \caption{The AE of the Green's function $\delta G_{\mathbf{k}=0}(\tau=0.1)$ for the one-dimensional Fermi-Hubbard model as a function of the lattice size $L$. The parameters are set to be $U=10$, $d=5$, and $\tau=0.1$. The blue dots represent the results obtained by the LVQC method, in which the time evolution operator for the $L\times1$ lattice model is prepared through the optimization of the local cost function $C_{\rm LHST}^f$ at the compilation size $\tilde{L}=2\times1$. The grey squares represent the results obtained by the Trotter decomposition of the time evolution operator for the $L\times1$ lattice model. } 
    \label{fig:1d-size}
\end{figure}
Figure~\ref{fig:1d-size} shows the AE $\delta G_{\mathbf{k}=0}(\tau)$ as a function of the size of the lattice $L$. 
Since the exact diagonalization of the full Hamiltonian $H^{(L)}$ is a computationally hard task when the system size $L$ is large, the exact value $G^{\rm R,exact}_{\mathbf{k}}$ is substituted by the Trotter decomposition with a sufficiently large depth $d=100$. 
We see that the AE $\delta G_{\mathbf{k}=0}(\tau)$ of the LVQC method is much smaller than that of the Trotter decomposition in a wide range of $L$. 
For example, the value of $\delta G_{\mathbf{k}=0}(\tau)$ of the LVQC method with $L=14$ is $2.21\times10^{-4}$, which is about 4 times smaller than the corresponding value of the Trotter decomposition $8.93\times10^{-4}$. 
In addition, we notice that the AE for both LVQC and Trotter decomposition are hardly altered with increasing the system size $L$. 
This can also be understood based on the similarity between the variational Hamiltonian ansatz of the LVQC method and the Trotter decomposition. 
Owing to the existence of the LR bound in the Fermi-Hubbard model, the Green's function $G^{\rm R}_{\mathbf{k}}(\tau)$ is considered as a local observable when the time $\tau$ is sufficiently small. 
The Trotter decomposition can simulate such a local observable with complexity independent of the system size~\cite{Childs-PRX-trotter}. 
This is consistent with the $L$ dependence of the AE $\delta G_{\mathbf{k}}(\tau)$ for the Trotter decomposition shown in Fig.~\ref{fig:1d-size} (grey diamonds).
The results for the LVQC method (blue circles) just inherit this property of the Trotter error owing to the similarity between
the variational Hamiltonian ansatz and the Trotter circuit. 
We also note that the $L$ dependence of the AE $\delta G_{\mathbf{k}}(\tau)$ for the LVQC method is consistent with Eq.~\eqref{eq:cost-lhst-lvqc} derived in Ref.~\cite{Mizuta-LVQC}, which states that the local cost function describing the local error of the ansatz hardly increases when the whole system size $L$ is increased while the compilation size $\tilde{L}$ is fixed. 

\subsection{Two-dimensional cases}\label{subsec:hubbard-2d}
\begin{figure}[tbp]
    \includegraphics[keepaspectratio, scale=0.63]{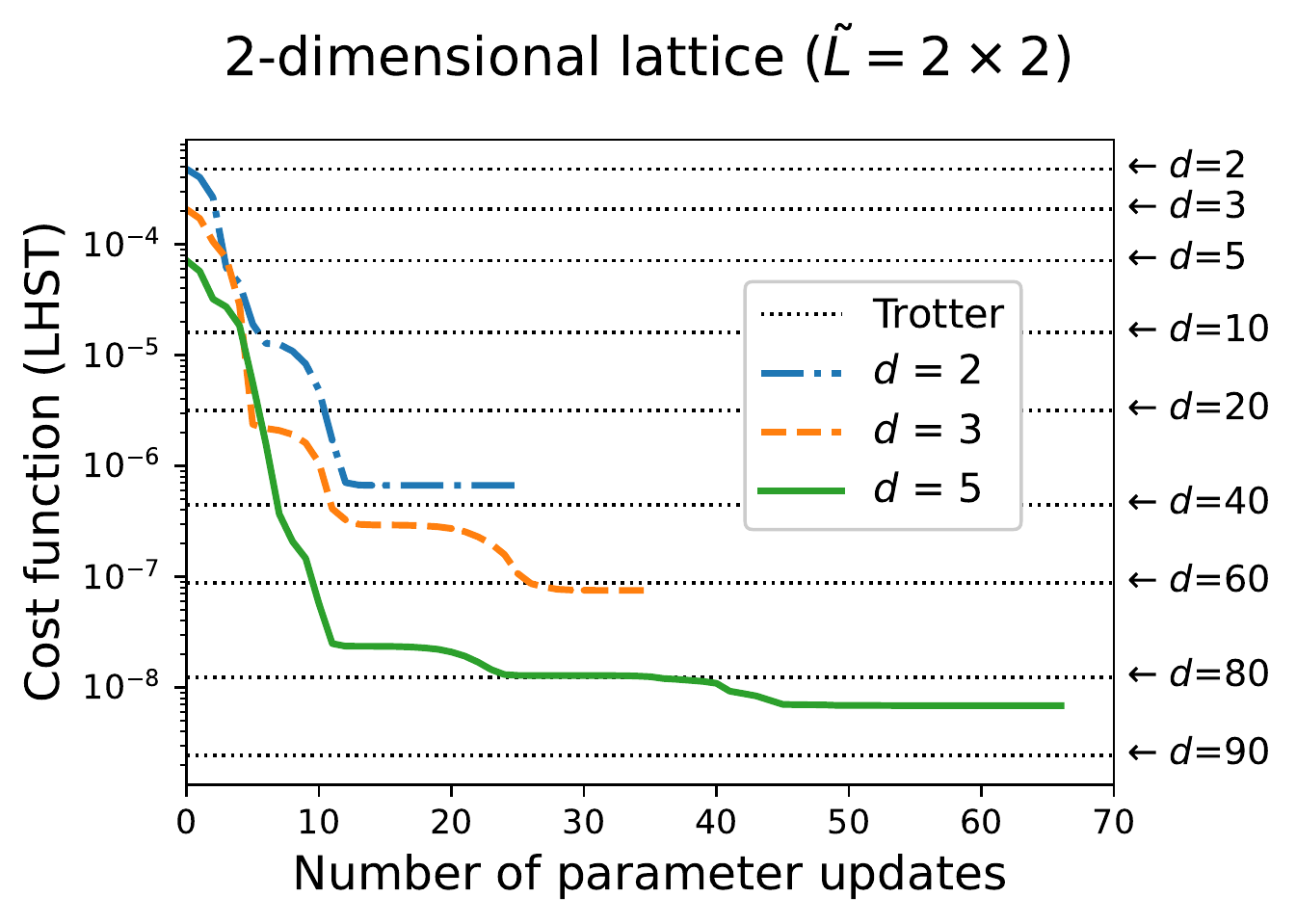}
    \caption{
    The history of optimization of the cost function $C_{\rm LHST}^f(U_{\rm 100}^{(\tilde{L})}(\tau),V_{d}^{(\tilde{L})}(\vec{\theta}))$ at the compilation size $\tilde{L}=2\times2$. The parameters are set to be $U=10$ and $\tau=0.1$. The blue dash-dotted line, orange dashed line, and green solid line represent the results for $d=2,3,5$, respectively.
    The black dotted lines represent the corresponding cost function for the Trotter decomposition $C_{\rm LHST}^f(U_{100}^{(\tilde{L})}(\tau),V_{d}^{(\tilde{L})}(\vec{\theta}_0))$ with various values of the depth $d$. } 
    \label{fig:2d-cost}
\end{figure}

Next, we show the numerical result for Case (II) and (III) shown in Figs.~\ref{fig:compile-size}(b) and (c).
Figure~\ref{fig:2d-cost} shows the history of the cost function $C_{\rm LHST}^f(U_{\rm 100}^{(\tilde{L})}(\tau),V_{d}^{(\tilde{L})}(\vec{\theta}))$ during the optimization at the compilation size $\tilde{L}=2\times 2$ and time $\tau=0.1$ for $d=2$ (blue dash-dotted line), $d=3$ (orange dashed line), and $d=5$ (green solid line). 
Similarly to the one-dimensional case, for each depth, $C_{\rm LHST}^f(U_{\rm 100}^{(\tilde{L})}(\tau),V_{d}^{(\tilde{L})}(\vec{\theta}))$ converges to the optimal value, which is significantly smaller than that of the corresponding Trotter decomposition. 
For example, the value of the local cost function $C_{\rm LHST}^f(U_{\rm 100}^{(\tilde{L})}(\tau),V_{d}^{(\tilde{L})}(\vec{\theta}))$ for $d=5$ is $6.85\times10^{-9}$, which is comparable to that of the depth-90 Trotter decomposition with $2.43\times10^{-9}$.
In addition, the optimized ansatz with $d=5$ has $\bar{F}(U_{\rm 100}^{(\tilde{L})}(\tau),V_{d}^{(\tilde{L})}(\vec{\theta}_{\rm opt})) = 1-2.40\times10^{-8}$, which is comparable to the average gate fidelity of the depth-80 Trotter circuit $\bar{F}(U_{\rm 100}^{(\tilde{L})}(\tau),V_{80}^{(\tilde{L})}(\vec{\theta}_{0}))=1-2.80\times10^{-8}$. 
These results indicate that the number of gates needed to accurately approximate the time evolution operator at the compilation size $\tilde{L}=2\times2$ is successfully reduced to about $5/90=1/18$ compared to the Trotter decomposition.

\begin{figure*}[tbp]
    \includegraphics[keepaspectratio, scale=0.39]{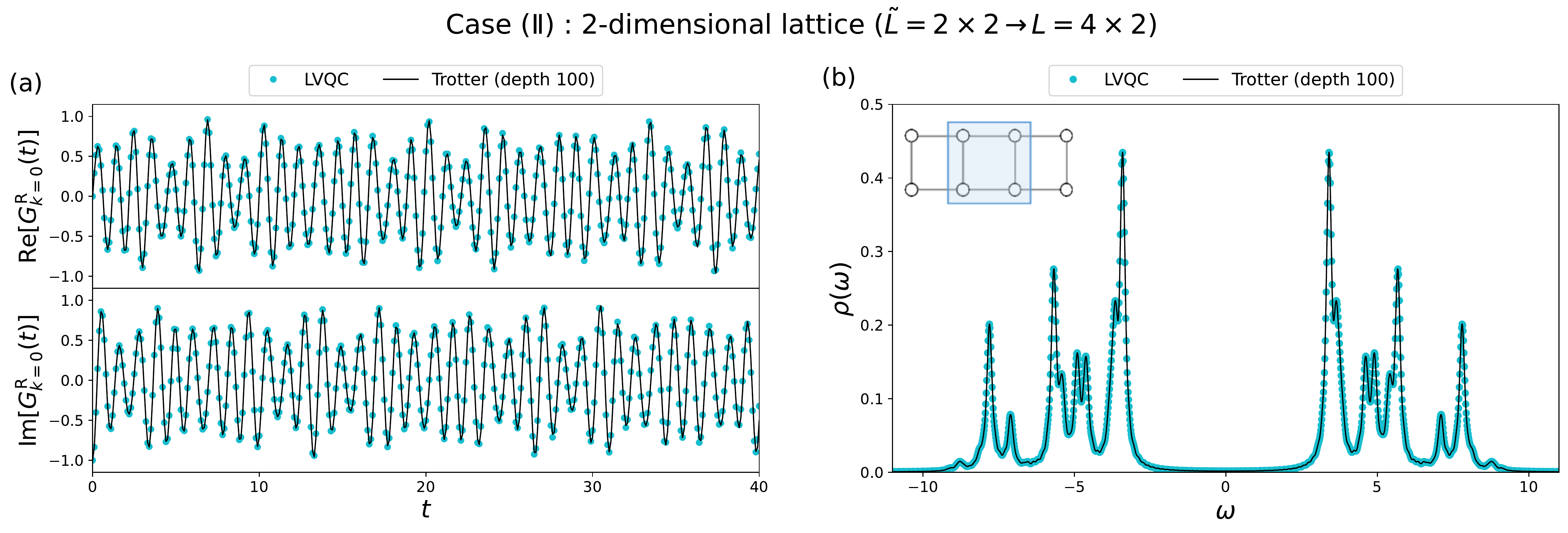}
    \caption{(a) The Green's function $G_{\mathbf{k}}^{\rm R}(t)$ at $\mathbf{k}=0$ and (b) the DOS $\rho(\omega)$ for the two-dimensional Fermi-Hubbard model on a $4\times 2$ lattice at $U=10$ and half-filling. 
    The cyan circles represent the results obtained by the LVQC algorithm which is performed through the minimization of the cost function $C_{\rm LHST}^f$ on a $2\times 2$ lattice with $d=5$ and $\tau=0.1$. The black lines represent the almost exact results obtained by the Trotter decomposition with the depth $d=100$. 
    The Green's function is calculated in the time domain $t\in [0,40]$ with the step $\tau=0.1$. We take $\eta=0.1$ for the calculation of the DOS.} 
    \label{fig:lvqc-2d}
\end{figure*}

\begin{figure*}[tbp]
    \includegraphics[keepaspectratio, scale=0.39]{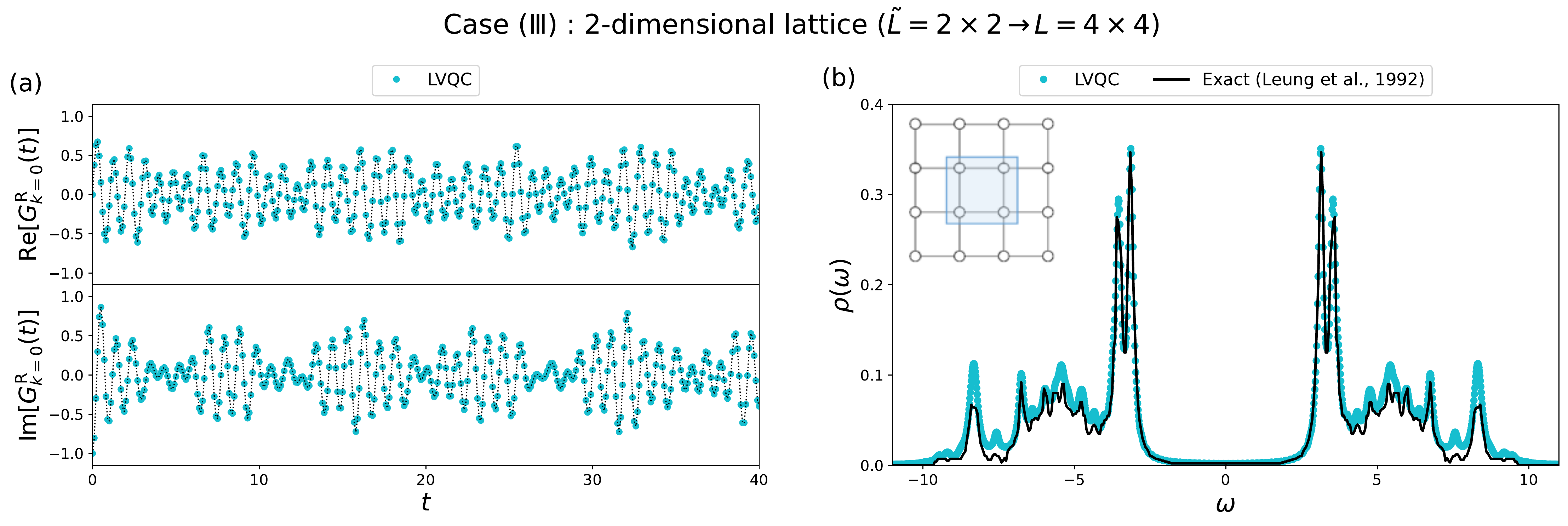}
    \caption{(a) The Green's function $G_{\mathbf{k}}^{\rm R}(t)$ at $\mathbf{k}=0$ and (b) the DOS $\rho(\omega)$ for the two-dimensional Fermi-Hubbard model on a $4\times 4$ lattice at $U=10$ and half-filling.
    The cyan circles represent the results obtained by the LVQC algorithm which is performed through the minimization of the cost function $C_{\rm LHST}^f$ on a $2\times 2$ lattice with $d=5$ and $\tau=0.1$. The black dotted line in panel (a) is for visibility.  
    The black line in panel (b) represents the exact value of the DOS taken from Ref.~\cite{4x4-exact}. 
    The Green's function is calculated in the time domain $t\in [0,40]$ with the step $\tau=0.1$. We take $\eta=0.1$ for the calculation of the DOS.} 
    \label{fig:lvqc-4x4}
\end{figure*}

The Green's function and DOS of the $4\times 2$ and $4\times 4$ lattice Fermi-Hubbard model calculated using the LVQC algorithm are shown in Figs.~\ref{fig:lvqc-2d} and \ref{fig:lvqc-4x4}, respectively.  
The LVQC algorithm is performed through the minimization of the cost function $C_{\rm LHST}^f(U_{\rm 100}^{(\tilde{L})}(\tau),V_{d}^{(\tilde{L})}(\vec{\theta}))$ at the compilation size $\tilde{L}=2\times 2$ and time $\tau=0.1$ with the ansatz depth $d=5$. 
In the simulation of the $4\times 4$ lattice model (Fig.~\ref{fig:lvqc-4x4}), we reduce the computational costs by using the symmetry of the Hamiltonian (see Appendix~\ref{append:symmetry}).
In Fig.~\ref{fig:lvqc-2d}, we compare the results of the LVQC algorithm with that of the Trotter decomposition with a sufficiently large depth $d=100$, since the exact calculation of the Green's function is a computationally hard task for the $4\times 2$ lattice Fermi-Hubbard model. 
The LVQC algorithm nicely reproduces the almost exact Green's function and DOS of the $4\times 2$ lattice Fermi-Hubbard model. 
We also ensure that the average gate fidelity of the optimized circuit $V^{(L)}_{d}(\vec{\theta}_{\rm opt})$ at $L=4\times2$ is better than the corresponding Trotter circuit. 
Specifically, the optimized ansatz extended to the whole lattice size $L=4\times2$ has $\bar{F}(U_{\rm 100}^{(L)}(\tau),V_{d}^{(L)}(\vec{\theta}_{\rm opt})) = 1-1.04\times10^{-7}$ for $d=5$, which is comparable to the average gate fidelity of the depth-80 Trotter circuit $\bar{F}(U_{\rm 100}^{(L)}(\tau),V_{80}^{(L)}(\vec{\theta}_{0}))=1-8.34\times10^{-8}$.
In Fig.~\ref{fig:lvqc-4x4}, we compare the results of the LVQC algorithm with the exact value shown in Ref.~\cite{4x4-exact}, in which the DOS is obtained by the Lanczos method. 
Note that the real-time Green's function is not computed in Ref.~\cite{4x4-exact} because the Lanczos method directly calculates the Green's function in the frequency domain. 
The LVQC algorithm nicely reproduces the overall peak structure of the exact DOS of the $4\times 4$ lattice Fermi-Hubbard model. 
Since the accuracy of the LVQC method for computing the Green's function is hardly altered with increasing the system size (see Fig.~\ref{fig:1d-size}), we expect that the LVQC method is also valid for accurately calculating the Green's function of the classically-intractable size of lattice more than $4\times 4$ lattice. 

\begin{figure*}[tbp]
    \includegraphics[keepaspectratio, scale=0.39]{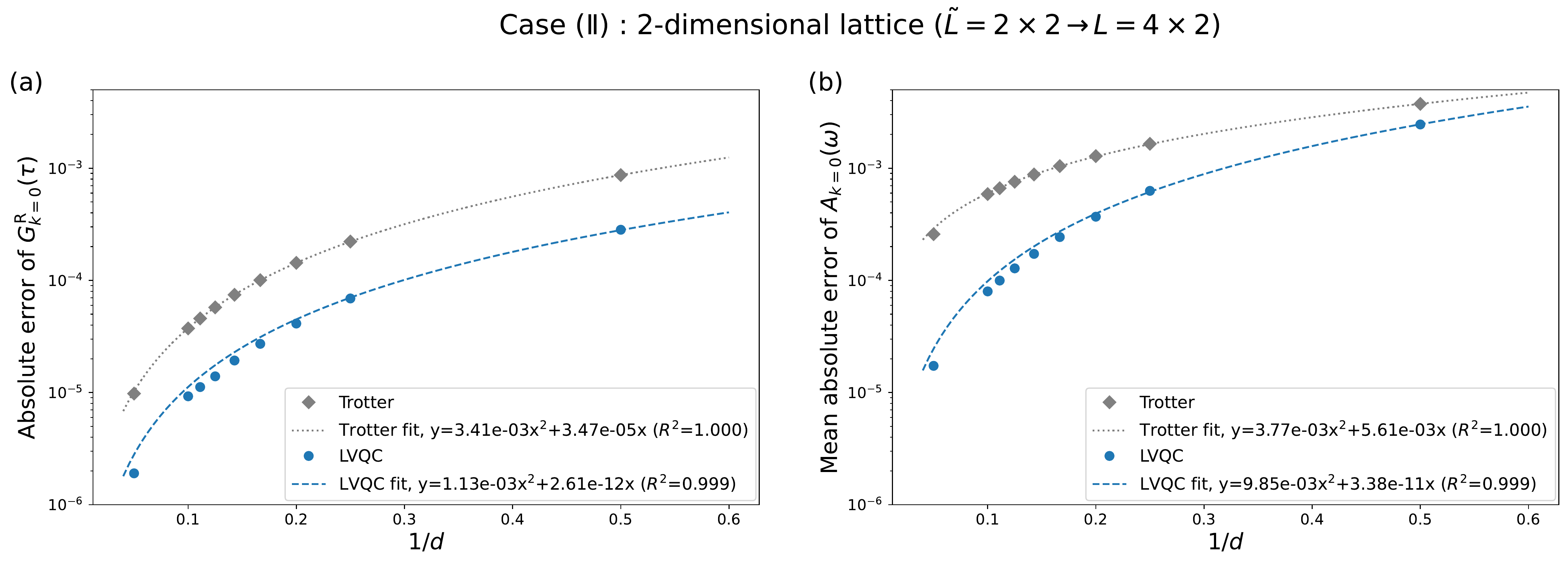}
    \caption{(a) The AE of the Green's function $\delta G_{\mathbf{k}=0}(\tau=0.1)$ and (b) the MAE of the spectral function $\delta A_{\mathbf{k}=0}$ for the two-dimensional Fermi-Hubbard model on a $4\times 2$ lattice at $U=10$ and half-filling. 
    The spectral function is calculated by the Fourier transformation of the Green's function $G_{\mathbf{k}}^{\rm R}(t)$ at $t\in[0,50]$ with the step $\tau=0.1$. 
    The broadening is taken as $\eta=0.1$. The MAE is calculated by setting $\omega_{\rm c}=15$ and $N_\omega=1000$. 
    The blue dots represent the results obtained by the LVQC method with the compilation size $\tilde{L} = 2\times 2$ and time $\tau=0.1$. The optimization is performed for the local cost function $C_{\rm LHST}^f$.
    The grey diamonds represent the results obtained by the Trotter decomposition of the time evolution operator for the $4\times 2$ lattice. 
    The blue dashed (grey dotted) line represents the fitting of the data of the LVQC (Trotter decomposition) by a quadratic function $y=\alpha x^2+\beta x$ based on the non-linear least squares. The values of the coefficients $\alpha$, $\beta$, and the coefficient of determination $R^2$ are shown in the legend. } 
    \label{fig:2d-depth}
\end{figure*}

Figure~\ref{fig:2d-depth} shows the dependence of the accuracy of the LVQC method on the depth of the ansatz $d$. 
We see that the AE $\delta G_{\mathbf{k}}^{\rm R}(\tau)$ (Eq.~\eqref{eq:error-gf}) and MAE $\delta A_{\mathbf{k}}$ (Eq.~\eqref{eq:error-spectral}) at $\mathbf{k}=0$ of the LVQC method is much smaller than that of the Trotter decomposition in a wide range of $d$. 
For example, the value of $\delta G_{\mathbf{k}=0}(\tau)$ of the LVQC method with $d=5$ is $4.12\times10^{-5}$, which is about 3 times smaller than the corresponding value of the Trotter decomposition $1.43\times10^{-4}$.
The value of $\delta A_{\mathbf{k}=0}$ of the LVQC method with $d=5$ is $3.70\times10^{-4}$, which is about 3 times smaller than the corresponding value of the Trotter decomposition $1.28\times10^{-3}$.
In addition, the $1/d$ dependence of the AE $\delta G_{\mathbf{k}}^{\rm R}(\tau)$ and MAE $\delta A_{\mathbf{k}}$ are well fitted in the form of $\alpha/d^2+\beta/d$ for both LVQC and Trotter decomposition. 
This result is the same as the one-dimensional case (Sec.~\ref{subsec:hubbard-1d}) and interpreted as that the LVQC method inherits the scaling property of the Trotter error because of the similarity between the variational Hamiltonian ansatz and the Trotter circuit. 

\section{Resource estimation} \label{sec:ResourceEstimation}
In this section, 
we discuss the feasibility of the LVQC method from the viewpoint of the computational resource.
We estimate the gate count and the number of shots needed to compute the spectral function for a given precision.
Unless otherwise noted, we consider the Fermi-Hubbard model on a two-dimensional
square lattice under a periodic boundary condition at half-filling.
We also assume that the total number of sites $L$ is even.

Let the spectral function should be computed within the precision $\varepsilon$. 
More precisely, MAE of the spectral function (Eq.~\eqref{eq:error-spectral}) is bounded as $\delta A_{\mathbf{k}} \leq \varepsilon$.
We also require the spectral function to have enough resolution for any frequency.
As discussed in Ref.~\cite{Endo-NISQ-2020}, 
this requirement is satisfied by setting the times step $\Delta t$ and the total integration time $T$ as 
\begin{align}
    \Delta t \leq \frac{\pi}{\Delta E_\text{max}}, \quad 
    T \geq \frac{2\pi}{\Delta E_\text{min}},
\end{align}
where $\Delta E_\text{max (min)}$ is the largest (smallest) energy difference between any two energy levels.
We set these values so that $N_t = T/\Delta t$ is an integer.
Then, the Fourier transform of $G^\text{R}_{\mathbf{k}\sigma}(t)$ is discretized as Eq.~\eqref{eq:green-fourier-sum}.
Let $\varepsilon_\text{disc}$ be the discretization error. 

Next, we consider AE~\eqref{eq:error-gf} of the Green's function.
In the LVQC method, 
the error consists of 
\begin{align}
    \delta G_\mathbf{k}(t)
    =
    \varepsilon'(t) + \varepsilon_\text{stat},
\end{align}
where $\varepsilon'(t)$ and $\varepsilon_\text{stat}$ denote the error coming from the infidelity of the time evolution operator and the statistical error, respectively.
Although $\varepsilon'(t)$ depends on the details of the ansatz, qualitatively, it is expected to behave as follows.
In the LVQC method,
the variational operator $V$ is compiled for a finite time-interval $\tau$ that is much smaller than $T$ in general.
Therefore, the time evolution operator at $t = N_t\tau$ behaves as $U(t) \simeq (V+\Delta V)^{N_t} \simeq V+N_t\Delta V$, where $\Delta V$ is the error of the variational operator.
This means that the error accumulates linearly in time.
From this observation,
we assume that $\varepsilon'(t)$ is a linear function of $t$ as 
\begin{align}
    \varepsilon'(t) \simeq kt,
    \label{eq:epsilonprime}
\end{align}
for some $k>0$.
In particular, $\varepsilon'(t)$ is upper-bounded by $\varepsilon'(T)$, $\varepsilon'(t) \leq \varepsilon'(T)$.
Putting these quantities into the definition of the spectral function and its MAE, 
we obtain
\begin{align}
    \frac{1}{\pi} 
    (
        (\varepsilon'(T)+\varepsilon_\text{stat})N_t\Delta t
        + 
        \varepsilon_\text{disc}
    )
    <
    \varepsilon.
\end{align}
Below, we ignore $\varepsilon_\text{disc}$ since it is the order of $(\Delta t)^2$.
We find that the error of Green's function should satisfy
\begin{align}
    \varepsilon'(T), \ \varepsilon_\text{stat} \lesssim \frac{\pi \varepsilon}{T}.
    \label{eq: precision requirement}
\end{align}
These inequalities for $\varepsilon'(T)$ and $\varepsilon_\text{stat}$ provide lower bounds of the gate count and the number of shots required, respectively.

First, we discuss how the inequality for $\varepsilon'(T)$ relates to the gate count.
While we have given a general argument so far,  we make several assumptions for $\varepsilon'(t)$ based on our numerical simulation.
The coefficient $k$ in Eq.~\eqref{eq:epsilonprime} is related to the depth of the ansatz through the results shown in Fig.~\ref{fig:2d-depth} (a) that relates $\varepsilon'(t)$ to $d$ at $t=\tau=0.1$.
Thus, we obtain
\begin{align}
    k = \frac{10\alpha}{d^2} + \frac{10\beta}{d},
\end{align}
where $\alpha=1.13 \times 10^{-3}$ and $\beta=2.61 \times 10^{-12}$.
From the condition Eq.~\eqref{eq: precision requirement}, 
the depth of the ansatz should satisfy
\begin{align}
    d 
    \gtrsim
    \frac{2\alpha}{\beta}
    \frac{1}{
       \sqrt{1 + 2\pi \alpha \varepsilon/( 5\beta^2 T^2)}
       -
       1
    }.
    \label{eq: requirement for d}
\end{align}
We note that this requirement for $d$ is quite conservative.
Indeed, Eq.~\eqref{eq: requirement for d} becomes $d \gtrsim 95$ for $\varepsilon=10^{-3}, T=50$ while Fig.~\ref{fig:2d-depth} (b) indicates that $d = O(1)$ is enough for the similar accuracy.

The depth of the ansatz directly relates to the complexity of the circuit to compute Green's function shown in Fig.~\ref{fig:circuit-gf}.
Here, we regard the $R_z$ gate and CNOT gate as elementary gate sets and count the number of these gates.
The leading contribution is $V(\vec{\theta})$.
Recalling that a unitary operator whose form is $e^{-i\theta P}$ contains one $R_z$ gate and $2n$ CNOT gates if $P$ is an $n$-qubit Pauli string,
the total number of $R_z$ gate in $V_d^{(L)}$ is $13Ld$ and that of CNOT gate is $4(4L^{3/2}+7L-4L^{1/2})d$. The derivation of this result is shown in Appendix~\ref{append:gate-count}.
The rest contribution comes from controlled $P_a^{(n)}$ and $P_b^{(m)}$ gates.
The number of CNOT gates in these controlled gates is equal to the length of $P_a^{(n)}$ and $P_b^{(m)}$ as a Pauli string. Without loss of generality, we can set $a=1$ due to the translation symmetry.
Thanks to the point group symmetry, the most distant site from $a=1$ is the point at the lower right corner of the square with side length $L/2+1$ that is $b=(L+L^{1/2}+2)/2$ when $L/2$ is even and $b=(L+L^{1/2})/2$ when $L/2$ is odd. (See Appendix~\ref{append:symmetry-hubbard} for the spatial symmetry of the Fermi-Hubbard model on a square lattice.)
We adopt the former case below.
Therefore, we obtain
\begin{align}
    N_{R_z} &= 13LN_td, \label{eq: gate count 1q} \\
    \begin{split}
    N_\text{CNOT} &= 
    4(4L^{3/2}+7L-4L^{1/2})N_td \\
    &\qquad+ \frac{L+L^{1/2}+2}{2}.
    \label{eq: gate count}
    \end{split}
\end{align}

The number of shots required per an element of Green's function $N_\text{shots}$ at a given time is straightforwardly estimated as 
\begin{align}
    N_\text{shot} &\sim \frac{1}{(\pi \varepsilon/T)^2},
\end{align}
due to the relation $\varepsilon_\text{stat} \propto 1/\sqrt{N_\text{shots}}$.
To compute the spectral function, we need $(\sqrt{L}+2)(\sqrt{L}+4)/4$ independent elements of Green's function (see Appendix~\ref{append:symmetry}) and $N_t$-times executions of the quantum circuit.
Thus, the total number of shots in the whole process reads
\begin{align}
    N_\text{tot}
    \sim
    \frac{N_t(L+6L^{1/2}+8)}{4(\pi \varepsilon/T)^2}.
    \label{eq: shot count}
\end{align}

For instance,
let us consider the LVQC method for an $L = 20 \times 20$ lattice, and say $\varepsilon =0.01$.
We also require the same time resolution and interval as the simulation on the $4 \times 2$ lattice, which are $\tau = 0.1$ and $T = 50$.
Putting these values into Eqs.~\eqref{eq: gate count 1q},~\eqref{eq: gate count}, and~\eqref{eq: shot count},
we find a conservative requirement of resources as
\begin{align}
    d &\gtrsim 30, \\
    N_{R_z} &\sim 7.8 \times 10^7, \\
    N_\text{CNOT} &\sim 2.1 \times 10^9, \\
   N_\text{tot} &\sim 1.7 \times 10^{11}.
\end{align}
The gate counts using the Trotter decomposition can be estimated in the same way by adopting $\alpha=3.41\times 10^{-3}$ and $\beta=3.47\times 10^{-5}$ in Eq.~\eqref{eq: requirement for d}, and we get
\begin{align}           
    d^\text{Trotter} &\gtrsim 68, \\
    N_{R_z}^\text{Trotter} &\sim 1.7 \times 10^8, \\
    N_\text{CNOT}^\text{Trotter} &\sim 4.7 \times 10^9.
\end{align}
The number of shots is the same as in the LVQC method since it does not depend on $\alpha$ and $\beta$.
Our results suggest that the LVQC method is more practical than Trotter decomposition regarding gate counts.
We stress again that the estimation of the gate counts and the number of shots here is a very conservative one (upper bound) to ensure the accuracy of the calculated Green's function.

\section{Discussion and conclusion}\label{sec:Summary}
In this paper, we proposed an efficient method to compute the Green's function on quantum computers by utilizing the LVQC algorithm. 
In the LVQC algorithm, we first execute the local compilation procedure on small-scale quantum devices or classical simulators and then approximately implements the time evolution operator for a large-scale quantum many-body system by a shallow-depth circuit. 
The Green's function is computed through the measurements of the time evolution circuit prepared by the LVQC protocol. 
Since the LVQC algorithm is useful to reduce the computational cost needed to simulate quantum dynamics on a broad level of quantum computers from NISQ devices to FTQCs, our method 
will be valid to efficiently compute Green's function for large-scale quantum many-body systems in various stages of the quantum computing era. 
To see the validity of our LVQC-based method, we performed numerical simulations for the one- and two-dimensional Fermi-Hubbard model. 
We showed that the LVQC-based method nicely reproduces the exact Green's function and DOS for both one- and two-dimensional cases. 
In addition, we verified that our LVQC-based method can compute the Green's function more efficiently and accurately compared to a standard approach in which the time evolution operator is implemented using the Trotter decomposition. 
Specifically, by calculating the AE of the Green's function and the MAE of the spectral function, we showed that our LVQC-based method is more accurate than the same-depth Trotter decomposition in a wide range of parameter regimes. 
The formal estimation of the gate count also indicates that our LVQC-based method has a practical advantage against the Trotter decomposition.
Although we focused on the one-particle Green's function at zero temperature throughout this paper, our LVQC-based method can be straightforwardly extended to compute other quantities, such as Green's function at finite-temperature and linear response functions. 

We here discuss some remarks on our method and numerical results.
First, there is a limitation to the choice of the ansatz in our LVQC-based method. 
As mentioned in Sec.~\ref{sec:LVQC}, for the fermionic Hamiltonian, the LVQC algorithm works only when the ansatz circuit is constructed in a form that preserves the parity of the number of fermions. 
Note that the variational Hamiltonian ansatz~\eqref{eq:ansatz-fermi-hubbard} adopted in Sec.~\ref{sec:Numerical} satisfies this requirement. 
Equivalently, if we wish to compute the Green's function for the spin or bosonic Hamiltonian using the LVQC algorithm, we have to adopt a local ansatz that is constructed from only local gates. 
Therefore, we need to carefully choose the ansatz considering the nature of the target Hamiltonian. 
Second, we remark that our numerical results in Sec.~\ref{sec:Numerical} show that the LVQC algorithm nicely reproduces the exact results even though the compilation size $\tilde{L}$ is much smaller than the lower bound given by Eq.~\eqref{eq:compile-size}. 
Here, we estimate the lower bound of the compilation size for $d=5$ and $\tau=0.1$, which is the parameter set adopted in the numerical simulations in Figs.~\ref{fig:lvqc-1d}, \ref{fig:lvqc-2d}, and \ref{fig:lvqc-4x4}. 
Since the Fermi-Hubbard model~\eqref{eq:fermi-hubbard} is composed of the on-site interaction and nearest-neighbor hopping, the range of interaction is $d_{H}=1$. 
Although the value of velocity $v$ is unclear, we can expect that $v\tau$ is not so large for $\tau=0.1$ because $v=\mathcal{O}(1)$ in general. 
Hence, we can estimate the lower bound of the compilation size as $\tilde{L}>12$ for $d=5$ and $\tau=0.1$ from Eq.~\eqref{eq:compile-size}. 
On the other hand, the results for the one-dimensional (two-dimensional) model in Fig.~\ref{fig:lvqc-1d} (Figs.~\ref{fig:lvqc-2d} and \ref{fig:lvqc-4x4}) are obtained by setting $\tilde{L}=2$ ($\tilde{L}=4$), which is smaller than the estimated lower bound. 
Therefore, we can expect that Eq.~\eqref{eq:compile-size} may overestimate the lower bound of the proper compilation size. 
A more detailed analysis of the relation between the error and the compilation size in the LVQC algorithm is left for future work. 

We believe that our work will open a new avenue for simulating the Green's function for large-scale quantum many-body systems, and encourage future research toward the realization of quantum advantage for practical problems in condensed matter physics, quantum chemistry, and material science.  

\section*{Acknowledgement}
The authors are grateful to Kaoru Mizuta for helpful discussions. 
S.K. acknowledges Masatoshi Ishii for technical support in performing numerical simulations.

\appendix

\section{Fermion-to-qubit mapping and Majorana fermions}\label{append:Majorana}
In this appendix, we show that one can choose to describe a fermion-to-qubit mapping in the following form:
\begin{align}
  \begin{split}
    c_{a} &\mapsto \frac{1}{2}\left(P_{a}^{(1)} + iP_{a}^{(2)} \right),\\
    c_{a}^{\dag} &\mapsto \frac{1}{2}\left(P_{a}^{(1)} -i P_{a}^{(2)} \right),\label{eq:qubit-map2}
  \end{split}
\end{align}
where $P_a^{(1)}$ and $P_a^{(2)}$ are Pauli operators satisfying $\{ P_{a}^{(n)}, P_{b}^{(m)} \} = 2\delta_{ab}\delta_{nm}$. 
Equation~\eqref{eq:qubit-map2} is a special case of Eq.~\eqref{eq:qubit-map}. 
To derive Eq.~\eqref{eq:qubit-map2}, we describe the fermionic operators in terms of Majorana fermions, 
\begin{align}
  \begin{split}
      c_{a} &= \frac{1}{2}(\gamma_{a}^{(1)} + i\gamma_{a}^{(2)}), \\
      c_{a}^{\dag} &= \frac{1}{2}(\gamma_{a}^{(1)} - i\gamma_{a}^{(2)}),
  \end{split} \label{eq:majorana}
\end{align}
where $\gamma_{a}^{(n)}$ $(n=1,2)$ are the Majorana operators. 
Since the Majorana operators $\gamma_{a}^{(n)}$ obey a Clifford algebra as $(\gamma_{a}^{(n)})^{\dag}=\gamma_{a}^{(n)}$ and $\{ \gamma_{a}^{(n)}, \gamma_{b}^{(m)} \} = 2\delta_{ab}\delta_{nm}$, they can be represented by the Pauli matrices $P_{a}^{(n)}$~\cite{majorana-qubitmap-2022}. 
Thus, by setting $\gamma_{a}^{(n)} \mapsto P_{a}^{(n)}$ in Eq.~\eqref{eq:majorana}, we obtain Eq.~\eqref{eq:qubit-map2}. 
Indeed, standard techniques such as Jordan-Wigner encoding~\cite{Jordan1928}, Bravyi-Kitaev encoding~\cite{bravyi-kitaev-2002} and parity encoding~\cite{seeley2012bravyi} take the form of Eq.~\eqref{eq:qubit-map2}.

\section{Symmetry-based reduction of measurements to compute the Green's function }\label{append:symmetry}
As discussed in Sec.~\ref{subsec:Green-quantum}, we can calculate the real-time Green's function by using the quantum circuit shown in Fig.~\ref{fig:circuit-gf}, whose measurement outcome gives the approximate value of $K_{a,b}^{(n,m)}(t)$ defined by Eq.~\eqref{eq:gf-pauli-kernel}. 
When the total number of all possible patterns of $K_{a,b}^{(n,m)}$ with respect to $(a,b; n,m)$ is $N_{\rm circ}$, we need to perform such measurements for $N_{\rm circ}$ distinct quantum circuits to obtain all components of the Green's function. 
In this appendix, we show that not all components of $K_{a,b}^{(n,m)}(t)$ are independent by considering the symmetry of the target system. 
This indicates that we can compute all the components of the Green's function by measuring some distinct quantum circuits less than $N_{\rm circ}$. 
Note that this approach is rigorously applicable only when the ansatz $V(\vec{\theta})$ commutes with the corresponding symmetry operations of the original fermionic Hamiltonian $H$. 


\subsection{General theory}
Here, we demonstrate how a component of $K_{a,b}^{(n,m)}(t)$ is connected to other components by symmetry.

\subsubsection{U(1) symmetry}
First, we consider the U(1) symmetry, which leads to the conservation of the particle number of the system. 
We here assume that the fermion-to-qubit mapping is performed in the form of Eq.~\eqref{eq:qubit-map2}. 
Considering the inverse transformation of Eq.~\eqref{eq:qubit-map}, the measurement outcome $K_{a,b}^{(n,m)}(t)$ defined by Eq.~\eqref{eq:gf-pauli-kernel} can be rewritten in terms of fermionic operators as 
\begin{align}
  \begin{split}
      K_{a,b}^{(1,1)}(t) &= \mathrm{Re}\langle(c_a(t)+c_a^{\dag}(t))(c_b+c_b^{\dag})\rangle_0, \\
    K_{a,b}^{(2,2)}(t) &= -\mathrm{Re}\langle(c_a(t)-c_a^{\dag}(t))(c_b-c_b^{\dag})\rangle_0, \\
    K_{a,b}^{(1,2)}(t) &= -i\mathrm{Re}\langle(c_a(t)+c_a^{\dag}(t))(c_b-c_b^{\dag})\rangle_0, \\
    K_{a,b}^{(2,1)}(t) &= -i\mathrm{Re}\langle(c_a(t)-c_a^{\dag}(t))(c_b+c_b^{\dag})\rangle_0, 
  \end{split} \label{eq:K-fermion}
\end{align}
where $c_a^{(\dag)}(t)\equiv e^{iHt}c_a^{(\dag)}e^{-iHt}$ and $\langle \cdots \rangle_0 \equiv \bra{\psi_0}\cdots\ket{\psi_0}$. 
If the Hamiltonian $H$ preserves the U(1) symmetry, 
\begin{equation}
    \langle c_a(t)c_b\rangle_0=\langle c_a^{\dag}(t)c_b^{\dag}\rangle_0=0, \label{eq:U(1)-particle-number}
\end{equation}
is satisfied since the particle number of the system is preserved.
From Eqs.~\eqref{eq:K-fermion} and~\eqref{eq:U(1)-particle-number}, we obtain 
\begin{align}
  \begin{split}
    K_{a,b}^{(1,1)}(t) &= K_{a,b}^{(2,2)}(t) = \mathrm{Re}\langle c_a(t)c_b^{\dag}+c_a^{\dag}(t)c_b\rangle_0, \\
    K_{a,b}^{(1,2)}(t) &= -K_{a,b}^{(2,1)}(t) = \mathrm{Re}\langle c_a(t)c_b^{\dag}-c_a^{\dag}(t)c_b\rangle_0.
  \end{split} \label{eq:kernel-U(1)} 
\end{align}
Thus, the number of independent combinations of $(n,m)$ is not 4 but 2, e.g., $(1,1)$ and $(1,2)$. 
Note that a similar result is also obtained in Ref.~\cite{Bauer-PRX-2016}. 

\subsubsection{Other symmetries}
We can further reduce the number of independent components of  $K_{a,b}^{(n,m)}(t)$ by considering symmetries other than U(1) symmetry. 
Suppose the Hamiltonian $H$ is invariant under a symmetry operation $\hat{R}$ as
\begin{equation}
    [H, \hat{R}] = 0 . \label{eq:H-symmetry}
\end{equation}
Using Eqs.~\eqref{eq:kernel-U(1)} and~\eqref{eq:H-symmetry}, we obtain
\begin{align}
  \begin{split}
    &K_{a,b}^{(n,m)}(t) \\
    &= \mathrm{Re}\langle e^{iHt} (\hat{R} c_a \hat{R}^{-1}) e^{-iHt} (\hat{R} c_b^{\dag}\hat{R}^{-1})\rangle_0 \\
    & \quad + \zeta_{nm}\mathrm{Re}\langle e^{iHt} (\hat{R} c_a^{\dag} \hat{R}^{-1}) e^{-iHt} (\hat{R} c_b \hat{R}^{-1})\rangle_0, \label{eq:K-symmetry}
  \end{split}
\end{align}
where 
\begin{equation}
    \zeta_{nm} = 
    \begin{cases}
        +1 & \text{for $n=m$}, \\
        -1 & \text{for $n\neq m$}.
    \end{cases}
\end{equation}
In the following, we apply Eq.~\eqref{eq:K-symmetry} to specific symmetries in quantum many-body systems. 
\begin{description}
    \item[Time-reversal symmetry] Suppose that the fermionic operators are specified by a spin index $\sigma(=\uparrow,\downarrow)$, i.e., $a=\sigma$. 
    Then, time-reversal operation $\hat{\Theta}$ is defined as 
    \begin{align}
       \begin{split}
           \hat{\Theta}c_{\uparrow}\hat{\Theta}^{-1}&=c_{\downarrow},\quad \hat{\Theta}c_{\uparrow}^{\dag}\hat{\Theta}^{-1}=c_{\downarrow}^{\dag}, \\
           \hat{\Theta}c_{\downarrow}\hat{\Theta}^{-1}&=-c_{\uparrow}, \quad \hat{\Theta}c_{\downarrow}^{\dag}\hat{\Theta}^{-1}=-c_{\uparrow}^{\dag}. \label{eq:time-reversal}
       \end{split}
    \end{align}
    For a time-reversal symmetric Hamiltonian (i.e., $[H,\hat{\Theta}]=0$), by setting $\hat{R}=\hat{\Theta}$ in Eq.~\eqref{eq:K-symmetry}, we obtain
    \begin{align}
        \begin{split}
            K_{\uparrow,\uparrow}^{(n,m)}(t) &= K_{\downarrow,\downarrow}^{(n,m)}(t), \\
            K_{\uparrow,\downarrow}^{(n,m)}(t) &= -K_{\downarrow,\uparrow}^{(n,m)}(t). \label{eq:K-TRS}
        \end{split}
    \end{align}
    Thus, the number of independent combinations of the spin indices is not 4 but 2, e.g., $(\uparrow,\uparrow)$ and $(\uparrow,\downarrow)$. 
    \item[Space group symmetry] 
    The spatial symmetry of periodic materials can be generally classified based on the space group $G$. 
    This means that the Hamiltonian is invariant under the space group operations $\hat{g}_{\ell} \in G$, i.e., $[H, \hat{g}_{\ell}]=0$.  
    Suppose that the transformation of the index $a(=1,2,\cdots,M)$ under a space group operation $\hat{g}_{\ell}$ is described by a permutation $\pi_{\ell}$; 
    \begin{equation}
        \pi_{\ell} = 
        \begin{pmatrix}
          1 & 2 & \cdots & M \\
          \pi_{\ell}(1) & \pi_{\ell}(2) & \cdots & \pi_{\ell}(M)
        \end{pmatrix}.
    \end{equation}
    Then, the fermionic operators are transformed under the space group operation $\hat{g}_{\ell}$ as
    \begin{align}
        \hat{g}_{\ell}c_{a}\hat{g}_{\ell}^{-1} = c_{\pi_{\ell}(a)}, \quad \hat{g}_{\ell}c_{a}^{\dag}\hat{g}_{\ell}^{-1} = c_{\pi_{\ell}(a)}^{\dag}. \label{eq:point-group}
    \end{align}
    From Eqs.~\eqref{eq:K-symmetry} and \eqref{eq:point-group}, we obtain
    \begin{equation}
        K_{a,b}^{(n,m)}(t) = K_{\pi_{\ell}(a),\pi_{\ell}(b)}^{(n,m)}(t). \label{eq:K-point-group}
    \end{equation}
    Note that the importance of considering the space group symmetry to reduce the resource for quantum simulation is also pointed out for symmetry-adopted VQE~\cite{SekiPRA2022}.
    \item[Particle-hole symmetry] Suppose the index $a$ denotes a site index on a lattice model. 
    When the lattice has a bipartite structure composed of sublattices $A$ and $B$, the particle-hole transformation $\hat{\Xi}$ is defined as 
    \begin{align}
        \hat{\Xi} c_{a} \hat{\Xi}^{-1} = \eta_{a}c_{a}^{\dag}, \label{eq:particle-hole-transfrom}
    \end{align}
    where 
    \begin{align}
        \eta_a = 
        \begin{cases}
            +1 & \text{for $a \in A$ sublattice}, \\
            -1 & \text{for $a \in B$ sublattice}.
        \end{cases}
    \end{align}
    For particle-hole symmetric systems, Eqs.~\eqref{eq:K-symmetry} and \eqref{eq:particle-hole-transfrom} leads to
    \begin{align}
        K_{a,b}^{(n,m)}(t) &= 
        \begin{cases}
          (1+\eta_a\eta_b)\mathrm{Re}\langle c_{a}(t)c_{b}^{\dag}\rangle_0 & \text{for $n=m$}, \\
          i(1-\eta_a\eta_b)\mathrm{Re}\langle c_{a}(t)c_{b}^{\dag}\rangle_0 & \text{for $n\neq m$}.
        \end{cases}
        \label{eq:K-PHS}
    \end{align}
    This indicates that only the diagonal components of $K_{a,b}^{(n,m)}(t)$ with respect to $(n,m)$ is nonzero when the sites $a$ and $b$ belong to same sublattices (i.e., $\eta_a\eta_b=1$). 
    On the other hand, only the off-diagonal components of $K_{a,b}^{(n,m)}(t)$ with respect to $(n,m)$ is nonzero when the sites $a$ and $b$ belong to different sublattices (i.e., $\eta_a\eta_b=-1$). 
\end{description} 

\subsection{Application to Fermi-Hubbard model}\label{append:symmetry-hubbard}
Next, we apply the above results to the Fermi-Hubbard model used in our numerical simulations in Sec.~\ref{sec:Numerical}, and demonstrate how to reduce the number of measurements in calculations of the Green's function. 
As seen from the Hamiltonian~\eqref{eq:fermi-hubbard}, the fermionic mode in the Fermi-Hubbard model is specified by the index of sites $i(=1,2,\cdots,L)$ and spin $\sigma(=\uparrow,\downarrow)$, i.e., $a=i\sigma$ and $M=2L$. 
By adopting the Jordan-Wigner encoding Eq.~\eqref{eq:jordan-wigner}, we obtain $N_{\rm circ}=4M^2=16L^2$. 
Now, we consider the symmetry of the Fermi-Hubbard model. 
From Eq.~\eqref{eq:fermi-hubbard}, we see that the Fermi-Hubbard model always preserves the U(1) symmetry and time-reversal symmetry. 
The particle-hole symmetry is preserved only at the half-filling (i.e., $\mu=U/2$)~\cite{arovas2022hubbard}. 
The detail of the space group symmetry depends on the lattice structure.
For a one-dimensional lattice, the Hamiltonian is invariant under point group operations of $C_2$. 
On the other hand, the Hamiltonian is invariant under point group operations of $C_{4v}$ ($C_{2v}$) for a two-dimensional square (rectangular) lattice with $L_x=L_y$ ($L_x\neq L_y$). 
In addition to such point group operations, translation symmetry is preserved under periodic boundary conditions. 

For concreteness, let us consider a two-dimensional square lattice under a periodic boundary condition. 
When the site number $L$ is even, the number of independent combinations of site indices $(i,j)$ under Eq.~\eqref{eq:point-group} is $(\sqrt{L}+2)(\sqrt{L}+4)/8$. 
By taking into account the results for U(1) symmetry~\eqref{eq:kernel-U(1)} and time-reversal symmetry~\eqref{eq:K-TRS}, the number of circuits needed to obtain all components of the Green's function is reduced to $4\times(\sqrt{L}+2)(\sqrt{L}+4)/8=(\sqrt{L}+2)(\sqrt{L}+4)/2$ from $16L^2$. 
At half-filling, this is further reduced to $(\sqrt{L}+2)(\sqrt{L}+4)/4$ based on Eq.~\eqref{eq:K-PHS}.
To summarize, in this case, the Green's function can be computed by performing measurements for $(\sqrt{L}+2)(\sqrt{L}+4)/4$ distinct quantum circuits. 
In the numerical simulation of $4\times4$ lattice Fermi-Hubbard model in Fig.~\ref{fig:lvqc-4x4}, we reduced the computational costs by calculating $K_{a,b}^{(n,m)}(t)$ only for such independent components. 
We justify such a procedure because the variational Hamiltonian ansatz~\eqref{eq:ansatz-fermi-hubbard} almost preserves the symmetry of the original Hamiltonian~\eqref{eq:fermi-hubbard}.
Although the space group operations can interchange the order of hopping terms $e^{i\theta_3^{(d)}P_t^{(r)}}$ in the ansatz~\eqref{eq:ansatz-fermi-hubbard}, we numerically confirmed that this has little or no effect on the results of the Green's function. 


\section{Local variational quantum compilation in fermionic systems \label{append:LVQC-fermion}}
The original paper of LVQC~\cite{Mizuta-LVQC} treated qubit (spin) systems, and the applicability of LVQC to fermionic systems was not explicitly discussed.
In this section, we describe a sketch of the proof of the LVQC theorem~\ref{thm:LVQC} for fermionic systems in Sec.~\ref{sec:LVQC}.

We consider a $L$-site and $\tL$-site fermionic systems with the Hamiltonians $H^{(L)}_{f,\mr{PBC}}$ and $H^{(\tL)}_{f,\mr{PBC}}$, respectively, described in Sec.~\ref{sec:LVQC}.
We assume the existence of the LR bound for $H^{(L)}_{f,\mr{PBC}}$ in the form of Eq.~\eqref{eq:LR-bound}, which is the case for general fermionic systems with finite-range interactions.
The LVQC theorem can be proved by the following inequalities:  
\begin{align}
 C_\mr{LHST}^{(\mu), f} (U^{(L)}_f, V^{(L)}_f) \leq C_\mr{LHST}^{(\mu), f} (U^{(L')}_f, V^{(L)}_f) + 2\varepsilon_\mr{LR}, \label{eq: LVQC ineq 1} \\ 
 C_\mr{LHST}^{(\mu), f} (U^{(L')}_f, V^{(L)}_f) = C_\mr{LHST}^{(\mu), f} (U^{(L')}_f, V^{(\tL)}_f), \label{eq: LVQC ineq 2} 
\end{align}
for $\mu = (L'/2, \uparrow), (L'/2, \downarrow)$.
Here, $L' \, (\leq \tL < L)$ is defined through a tunable parameter $l_0$ as
\begin{equation}
    L' := 2 (l_0 + d_H + v \tau),
\end{equation}
where $d_H$ is the range of the interactions and $v$ is the velocity in the LR bound~[Eq.~\eqref{eq:LR-bound}].
$\epsilon_\mr{LR}$ is $\order{e^{-l_0/\xi}}$, and $U^{(L')}_f = e^{-iH^{(L')}_{f,\mr{PBC}}\tau}$ is the time evolution operator of the translationally-invariant Hamiltonian defined on the size-$L'$ system $\Lambda^{(L')}$~[Eq.~\eqref{eq: def of lamnda_tN}].
We omit the arguments $\tau$ and $\vec{\theta}$ of $U^{(L)}_f, U^{(\tL)}_f$ and $V^{(L)}_f, V^{(\tL)}_f$ respectively for brevity.
In the following, we sketch the proof of these two inequalities in order.

\subsection{Derivation of \eqref{eq: LVQC ineq 1}}
First, we show the inequality~\eqref{eq: LVQC ineq 1}, which says that we can replace the time evolution operator $U^{(L)}_f$ with $U^{(L')}_f$.
By following Appendix C of Ref.~\cite{Else2020}, one can transform the LR bound~\eqref{eq:LR-bound} into 
\begin{align}
 \left\| (U^{(L)}_f)^\dag F_{j_\mu} U^{(L)}_f -  (U^{(L')}_f)^\dag F_{j_\mu} U^{(L')}_f \right\| \leq  \varepsilon_{\mr{LR}}, \\
 \varepsilon_{\mr{LR}} = C' \int_{L'/2 - d_H}^{\infty} \, dx e^{ -(x-v\tau)/\xi} = e^{-\order{l_0/\xi}},
\end{align}
where $j_\mu \,(=L'/2)$ is the site on which the mode $\mu$ is defined, $F_{j_\mu}$ is the fermionic operators acting on the site $j$,
and $C'$ is some constant.
We note that the derivation of the above inequality in Ref.~\cite{Else2020} was for bosonic (spin) systems. Still, it also holds for fermionic systems because the derivation relied on properties not specific to bosons, such as the triangle inequality of the operator norm.

Another important observation to derive Eq.~\eqref{eq: LVQC ineq 1} is that the LHST cost function $C^{(\mu),f}_\mr{LHST}(U_f, V_f)$ [Eq.~\eqref{eq: def C_LHST_mu}] can be expressed by the expectation value of the projection operator,
\begin{equation}
\Pi_\mu = \ket{\Phi^f_{+,\mu}} \bra{\Phi^f_{+,\mu}},
\end{equation}
for the state $(U_f \otimes V_f^*) \ket{\Phi^f_{+}}$ in the doubled system.
Then, the difference between the cost function is
\begin{align*}
 & \left| C^{(\mu),f}_\mr{LHST}(U_f^{(L)}, V_f^{(L)}) - C^{(\mu),f}_\mr{LHST}(U_f^{(L')}, V_f^{(L)}) \right|  \\
 = & \left| \ev{(U_f^{(L)} \otimes V_f^{(L)*})^\dag \Pi_\mu (U_f^{(L)} \otimes V_f^{(L)*})}{\Phi^f_+} - \right. \\
 & \left. \ev{(U_f^{(L')} \otimes V_f^{(L)*})^\dag \Pi_\mu (U_f^{(L')} \otimes V_f^{(L)*})}{\Phi^f_+} \right|.
\end{align*}
From the definition of $\ket{\Phi_{+,\mu}^f}$ [Eq.~\eqref{eq: def bell pair mu}], we can show 
\begin{align*}
 \Pi_\mu &= \frac{1}{2}( (1-n_{A_\mu})(1-n_{B_\mu}) + a_{A_\mu}^\dag a_{B_\mu}^\dag + a_{B_\mu} a_{A_\mu} + n_{A_\mu} n_{B_\mu}) \\
&= \frac{1}{2} \sum_{i=1}^4 \mathcal{F}^A_i \mathcal{F}^B_i
\end{align*}
where $n_j = c_j^\dag c_j$ and $\mathcal{F}_i^{A(B)}$ is the fermionic operator acting on the system $A (B)$ with a unit operator norm $\| \mathcal{F}_i^{A(B)} \| = 1$.
Therefore, we obtain
\begin{align*}
 & \left| C^{(\mu),f}_\mr{LHST}(U_f^{(L)}, V_f^{(L)}) - C^{(\mu),f}_\mr{LHST}(U_f^{(L')}, V_f^{(L)}) \right|  \\
\leq & \frac{1}{2} \sum_{i=1}^4  \left\| (U^{(L)}_f)^\dag \mathcal{F}_i^A U^{(L)}_f -  (U^{(L')}_f)^\dag \mathcal{F}_i^A U^{(L')}_f \right\| \times \\
& \left\| (V^{(L)}_f)^\dag \mathcal{F}_i^B V^{(L)}_f \right\| \\
& \leq 2 \varepsilon_\mr{LR},
\end{align*}
which is Eq.~\eqref{eq: LVQC ineq 1}. 

\subsection{Derivation of \eqref{eq: LVQC ineq 2}}
Derivation of the inequality~\eqref{eq: LVQC ineq 2} is completely parallel to the original discussion of LVQC~\cite{Mizuta-LVQC}, so we describe it briefly with remarking the points specific to fermionic systems.
Assuming the form of the ansatz~[Eqs.~\eqref{eq: ansatz for N} and \eqref{eq: ansatz for tildeN}], we can show that some parts of the fermionic gates in the ansatz $V^{(L)}_f$ are canceled out in the calculation of the LHST cost function,
\begin{align*}
 & C^{(\mu),f}_\mr{LHST}(U_f^{(L')}, V_f^{(L)}) \\
= & \ev{(U_f^{(L')} \otimes V_f^{(L)*})^\dag \Pi_\mu (U_f^{(L')} \otimes V_f^{(L)*})}{\Phi^f_+}.
\end{align*}
Considering that the $\Pi_\mu$ is a sum of the local fermionic operators acting on the site $j_\mu$ in the system $A$ and $B$, the fermionic gates outside the cause cone of $\Pi_\mu$ are canceled out because of the unitary of the gates.
Moreover, for the fermion Bell pair defined in Eq.~\eqref{eq: def bell pair},
\begin{equation}
 I \otimes V_f \ket{\Phi^f_+} = V_f^* \otimes I \ket{\Phi^f_+}
\end{equation}
holds for any fermionic operators $V_f$.
We can ``move" the fermionic gate in the ansatz on the system $A$ to the system B,
and the fermionic gates outside the causal cone of $U^{(L')}_f$ are canceled out (see Fig.~4 of Ref.~\cite{Mizuta-LVQC} for details).
Because of these two cancellations, we can show that the inequality~\eqref{eq: LVQC ineq 2} holds for 
\begin{equation}
  \tL \geq \frac{L'}{2} + 2d + 1 = l_0 + d_H + v\tau + 2d + 1,
\end{equation}
and $4d > \tL$, by the exactly same discussion with Ref.~\cite{Mizuta-LVQC}.

\subsection{Derivation of LVQC theorem for fermions}
With two inequalities~\eqref{eq: LVQC ineq 1} and \eqref{eq: LVQC ineq 2}, one can show the LVQC theorem for fermionic systems.
The translational invariance of the system results in
\begin{align*}
   & C^{f}_\mr{LHST}(U_f^{(L)}, V_f^{(L)}) \\
 =  & \frac{1}{2} \left( C^{(j,\uparrow),f}_\mr{LHST}(U_f^{(L)}, V_f^{(L)}) + C^{(j,\downarrow),f}_\mr{LHST}(U_f^{(L)}, V_f^{(L)})\right)
\end{align*}
for any site $j$.
For $\mu = (L'/2, \uparrow), (L'/2, \downarrow)$, the inequalities~\eqref{eq: LVQC ineq 1} and \eqref{eq: LVQC ineq 2} are utilized to show
\begin{align*}
  C^{(\mu),f}_\mr{LHST}(U_f^{(L)}, V_f^{(L)})  
  &\leq C^{(\mu),f}_\mr{LHST}(U_f^{(L')}, V_f^{(L)}) + 2\varepsilon_\mr{LR} \\
  &= C^{(\mu),f}_\mr{LHST}(U_f^{(L')}, V_f^{(\tL)}) + 2\varepsilon_\mr{LR}.
\end{align*}
It is also possible to show 
\begin{equation}
 C^{(\mu),f}_\mr{LHST}(U_f^{(L')}, V_f^{(\tL)}) \leq C^{(\mu),f}_\mr{LHST}(U_f^{(\tL)}, V_f^{(\tL)}) + 2\varepsilon_\mr{LR}
\end{equation}
by the same argument to prove Eq.~\eqref{eq: LVQC ineq 1},
so we obtain
\begin{equation}
 C^{(\mu),f}_\mr{LHST}(U_f^{(L)}, V_f^{(L)})  \leq C^{(\mu),f}_\mr{LHST}(U_f^{(\tL)}, V_f^{(\tL)}) + 4\varepsilon_\mr{LR},
\end{equation}
which means Eq.~\eqref{eq:cost-lhst-lvqc}.
For the HST cost function, we combine the inequality between the LHST and HST cost functions proved in Ref.~\cite{Khatri-QAQC}, $C^{(f)}_\mr{HST}(U_f, V_f) \leq  2L \cdot C^{(f)}_\mr{LHST}(U_f, V_f)$, and Eq.~\eqref{eq:cost-lhst-lvqc}.
Then Eq.~\eqref{eq:cost-hst-lvqc} holds.

We comment on the extension of the discussion in this section to the cases without translational invariance as well as general dimensions.
As shown in Ref.~\cite{Mizuta-LVQC}, the original LVQC theorem for spin systems can be extended to such cases,
and we expect that it still holds for fermionic systems.
This is because the proof did not rely on anything specific to spin systems as we presented in this section.

\section{Derivation of the gate count \label{append:gate-count}}
In this section, 
we derive the number of $R_z$ gates and CNOT gates in the quantum circuit that approximates the time evolution operator.
As we see in Eq.~\eqref{eq:ansatz-fermi-hubbard}, $V_d^{(L)}(\vec{\theta})$ consists of $e^{i\theta P_\mu^{(L)}}, e^{i\theta P_U^{(L)}}$ and $e^{i\theta P_{t_r}^{(L)}}$.
We count the number of elementary gates in these components one by one.
\begin{description}
\item[(1) $e^{i\theta P_\mu^{(L)}}$]
By $
    e^{i\theta P_\mu^{(L)}} 
    = 
    \prod_{i=1}^L
    \prod_{\sigma=\uparrow,\downarrow}
    e^{i\theta Z_{i_\sigma}}
$,
this operator contains $N_\text{spin}L$ $R_z$ gates and $2N_\text{spin}L$ CNOT gates, where $N_\text{spin} = 2$.

\item[(2) $e^{i\theta P_U^{(L)}}$]
By 
\begin{align}
    e^{i\theta P_U^{(L)}}
    =
    \prod_{i=1}^L
    e^{i\theta Z_{i_\uparrow}Z_{i_\downarrow}}
    e^{-i\theta Z_{i_\uparrow}}
    e^{-i\theta Z_{i_\downarrow}},
\end{align}
this operator contains $3L$ $R_z$ gates and $2(2+1+1)L = 8L$ CNOT gates.

\item[(3) $e^{i\theta P_{t_r}^{(L)}}$ $(r=1,3)$]
When $r=1,3$,
we can write
\begin{align}
    \prod_{r=2,4}e^{i\theta P_{t_r}^{(L)}} 
    =
    \prod_{\sigma=\uparrow,\downarrow}
    \prod_{\langle i,j \rangle_h}
    \prod_{P=X,Y}
    e^{i\theta P_{i_\sigma}Z_{i_\sigma+1}\cdots P_{j_\sigma}},
\end{align}
where $\prod_{\langle i,j \rangle_h}$ means taking the product with respect to all horizontal bonds.
For instance, in a $4 \times 4$ lattice, $(i,j)=(1,2),(2,3),(3,4),(4,1),(5,6), \cdots$.
There are $L$ horizontal bonds in total.
Thus, the above operator contains $2N_\text{spin}L$ $R_z$ gates.
The length of the Pauli string for a horizontal bond in the bulk region is two while that is $L_x$ at the boundary.
Thus, the number of CNOT gates is
\begin{align}
    4N_\text{spin}(2(L_x-1)L_y+L_xL_y)
    =
    4N_\text{spin}(3L-2L^{1/2}).
\end{align}

\item[(4) $e^{i\theta P_{t_r}^{(L)}}$ $(r=2,4)$]
When $r=2,4$,
we can write
\begin{align}
    \prod_{r=2,4}e^{i\theta P_{t_r}^{(L)}} 
    =
    \prod_{\sigma=\uparrow,\downarrow}
    \prod_{\langle i,j \rangle_v}
    \prod_{P=X,Y}
    e^{i\theta P_{i_\sigma}Z_{i_\sigma+1}\cdots P_{j_\sigma}},
\end{align}
where $\prod_{\langle i,j \rangle_v}$ means taking the product with respect to all vertical bonds.
There are $L$ vertical bonds in total.
Thus, the above operator contains $2N_\text{spin}L$ $R_z$ gates.
To count the number of CNOT gates, we clarify the length of $P_{i_\sigma}Z_{i_\sigma+1}\cdots P_{j_\sigma}$ as a Pauli string.
For a moment, we consider $\sigma=\uparrow$.
Each vertical bond is specified by a pair of sites whose indices are
\begin{align}
    (j-1)L_x + i, \quad
    (j+1)L_x -i + 1, 
\end{align}
where $1 \leq i \leq L_x$ and $1 \leq j \leq L_y-1$ in the bulk region and
\begin{align}
    j, \quad
    L-(j-1), 
\end{align}
where $1 \leq j \leq L_x$ at the boundary.
Thus, the length of the Pauli string corresponding to this pair is $2(L_x - i + 1)$ in the bulk region and $L+2-2j$ at the boundary.
Taking a summation of all vertical bonds, we get
\begin{align}
    &\sum_{j=1}^{L_y-1}
    \sum_{i=1}^{L_x}
    2(L_x-i+1)
    +
    \sum_{j=1}^{L_x}(L+2-2j) \\
    &=
    (L_y-1)L_x(L_x+1)
    +    
    L_x(L_xL_y-L_x+1) \\
    &=
    2L^{3/2}-L.
\end{align}
The same argument yields the same result for $\sigma=\downarrow$.
Therefore, 
the number of CNOT gates in $ \prod_{r=2,4}e^{i\theta P_{t_r}^{(L)}}$ is given by
\begin{align}
    4N_\text{spin}(2L^{3/2}-L).
\end{align}
\end{description}
Summarizing (1)-(4),
the total number of $R_z$ gate in $V_d^{(L)}(\vec{\theta})$ per layer is $13L$ and that of CNOT gate is $4(4L^{3/2}+7L-4L^{1/2})$.

%

\end{document}